\documentclass[12pt]{iopart}
\usepackage{amsfonts}
\usepackage{iopams}
\usepackage{amssymb}
\usepackage{setstack}
\usepackage{graphicx}

\begin{document}
\title{Scattering on two Aharonov-Bohm vortices with opposite fluxes}
\author{E  Bogomolny$^1$, S Mashkevich$^2$, and S Ouvry$^1$}
\address{$^1$ CNRS, Universit\'e Paris-Sud, UMR 8626\\
Laboratoire de Physique Th\'eorique et Mod\`eles Statistiques, 91405 Orsay,
France}
\address{$^2$ Schr\"{o}dinger, 120 West 45th St., New York, NY 10036, USA\\
and
Bogolyubov Institute for Theoretical Physics, Kiev 03680, Ukraine}
\eads{\mailto{eugene.bogomolny@lptms.u-psud.fr}, \mailto{mash@mashke.org}, \mailto{stephane.ouvry@lptms.u-psud.fr}}
\begin{abstract}
The scattering of an incident plane wave on two Aharonov-Bohm  vortices with  opposite fluxes is considered in detail.  The presence of the vortices  imposes non-trivial boundary conditions for the partial waves on a cut joining the  two vortices. These conditions result in an infinite system of equations for scattering  amplitudes between  incoming and outgoing partial waves, which can be   solved numerically. The main focus of the paper  is the  analytic determination of the scattering amplitude in two limits,   the small flux limit  and  the  limit of small vortex separation. In the latter limit the dominant contribution comes from the $S$-wave amplitude. Calculating it,  however, still requires  solving  an infinite system of equations, which   is achieved  by the Riemann-Hilbert method.  The results agree well with the numerical calculations.      
\end{abstract}
\maketitle

\section{Introduction}

The Aharonov-Bohm problem \cite{AB} certainly leads to one of the most fascinating  quantum mechanical   effects which has been experimentally tested \cite{tonomura}. 

Clearly the original model  as stated in the seminal 1959 paper \cite{AB} is by itself somehow an abstraction. Leaving aside the zero flux size limit issue and the  related question of the behaviour of  wave functions at the location of the vortex (see, e.g.,  \cite{comtet}) there is not such a physical system where the flux of a magnetic field can pierce a plane at a certain location without returning at some other place on the plane so that the total flux is  zero. In the Aharonov-Bohm paper this issue  is solved by assuming that the return flux is spread   far away on the plane (in fact at infinity)  so that its direct effect on the problem at hand can be neglected. This simplifying assumption is crucial  in rendering the model exactly solvable, and most of the research effort has been devoted to this situation (see, e.g., \cite{ruijsenaars}, \cite{book} and references therein).

We would like to address  here a more realistic  situation where the piercing flux returns at a finite distance.  More precisely, we  consider two vortices of opposite strength  piercing the plane at  distance  $2R$  one from the other.  Doing so, we move from the solvable one-vortex standard Aharonov-Bohm problem to  a two-vortex problem, which is  more difficult to address, with no explicit solution at disposal. 

A few many-vortex  Aharonov-Bohm problems have already been considered in the literature, in particular the scattering on vortices arranged on  a  lattice  \cite{4}  or vortices whose  locations are random \cite{desbois}.

The  problem of scattering of an incident plane wave on several vortices has mostly been discussed  by two  methods. In \cite{gu_1}--\cite{gu_3} a formal expansion of the scattering amplitude in a series of  Mathieu functions has been constructed and in  \cite{stovicek_1}--\cite{stovicek_3} a formal diagrammatic-like series has been proposed. So far these approaches have had a limited success and did not produce explicit formulas  except  when the distance between the vortices goes to infinity \cite{stovicek_3}. 

Here we consider the scattering problem on two vortices in the singular gauge  with non-trivial boundary conditions  along a cut connecting the vortices. To address this problem, we use   an approach  already put forward  in \cite {MMO} for studying the low-energy spectrum of a charged particle in an harmonic well coupled to two Aharonov-Bohm fluxes of different strength. In the plane a  discontinuity  has to materialize   on any branch cut  joining the two vortices due to the non-trivial Aharonov-Bohm phases accumulated by the charged particle moving around one or the other vortex.  The positions of the vortices are fixed at $R$ and $-R$ on the $x$-axis, so that the branch cut can be chosen to be the half-circle of radius $R$ centered at the origin  in the lower half-plane (of course the observables, here the modulus squared of the scattering amplitude, should not depend on the choice of a  branch cut). As here the total flux is zero, no phase is accumulated at infinity by the particle encircling both vortices.  The scattering amplitude is, by definition,  a series of  scattering  amplitudes between  incoming and outgoing Hankel partial waves. The non-trivial boundary conditions on the cut lead to an infinite system of equations for the expansion coefficients, which are solved numerically. The main part of the paper is devoted to an analytic investigation of two limiting cases. The first is the small flux limit (more precisely, the small $\alpha$ limit, where $\alpha$ is the dimensionless ratio of the flux to the quantum of  flux) and the second is  the small $R$ limit (in fact the small $kR$ limit, where $k$  is the scattering momentum). In both cases  the analytic expressions  obtained are in  good agreement with the numerical calculations. 

The plan of the paper is the following.  We start in Section~\ref{equations} by developing the formalism for the scattering of an incident plane wave making an angle $\theta$ with the $x$-axis. We illustrate in Section~\ref{nume} the physics at hand with  several numerical simulations in various cases of interest. Next, in Section~\ref{small_alpha} we consider the small $\alpha$ limit and find an exact lowest-order expression for the scattering amplitude. In Section~\ref{small_kR_case}  we consider the small $kR$ limit, which allows for some simplifications in the contribution of the relevant partial wave scattering amplitudes.  Then in Section~\ref{exact_solution} we transform the small $kR$ equations into a form which lets us  find an exact expression for the scattering amplitude  by the Riemann-Hilbert method. In Section~\ref{approach} we show how the same solution can be obtained and simplified  by applying the Riemann-Hilbert method directly to the problem. In Section~\ref{conclusion} we outline the main results. A few details of the calculations are given in the Appendix.

\section{General formalism}\label{equations}

\begin{figure}
\begin{center}
\includegraphics[width=.5\linewidth]{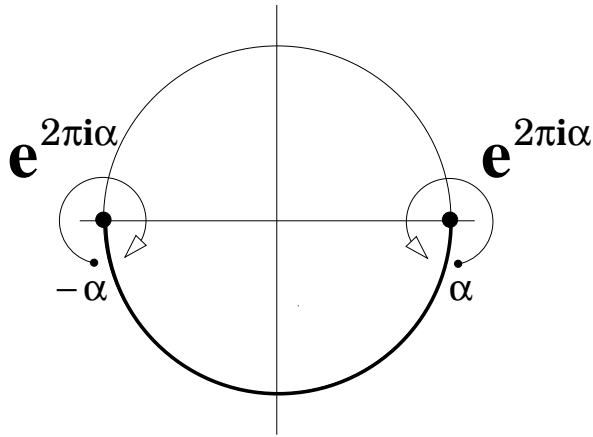}
\end{center}
\caption{The branch cut in the plane and the boundary conditions for two Aharonov-Bohm vortices with fluxes $\alpha$ and $-\alpha$. }
\label{fig_cut}
\end{figure}

We consider two Aharonov-Bohm vortices  on a plane at  points $\vec{R}$ and $-\vec{R}$, i.e., with polar coordinates $(R,0)$ and $(R,\pi)$ with  fluxes $\alpha$ and $-\alpha$ respectively (see figure~\ref{fig_cut}). We work in  the singular gauge, where the electromagnetic potential is removed but the wave function is defined  with a cut connecting the two vortices such that on the opposite branches of the cut the wave function and its normal derivative obey the relations 
\begin{equation}
\Psi^{(-)}(R,\phi)=f(\phi)\Psi^{(+)}(R,\phi)
\label{AB_function}
\end{equation}
and
\begin{equation}
\frac{\partial}{\partial r}\Psi^{(-)}(R,\phi)=f(\phi) \frac{\partial}{\partial r}\Psi^{(+)}(R,\phi).
\label{AB_derivative}
\end{equation}
Here and below superscripts $(+)$ and $(-)$ indicate functions  outside and inside the circle, respectively, and $f(\phi)$ describes the necessary jump along the cut. In the geometry of figure~\ref{fig_cut} it can be written in various equivalent forms
\begin{equation}
\fl
f(\phi)=\left \{ \begin{array}{cc} \mathrm{e}^{2\pi \mathrm{i} \alpha},& \phi  \in I\\1, & \phi  \notin I \end{array}\right .=1+
\left \{ \begin{array}{cc} (\mathrm{e}^{2\pi \mathrm{i} \alpha}-1),& \phi  \in I\\0, & \phi   \notin I \end{array}\right .
= \mathrm{e}^{\pi \mathrm{i} \alpha}\left \{ \begin{array}{cc} 
\mathrm{e}^{\pi \mathrm{i} \alpha},& \phi  \in I\\ \mathrm{e}^{-\pi \mathrm{i} \alpha}, & \phi  \notin I \end{array}\right. 
\label{phi}
\end{equation} 
where $I=[\pi\ldots 2\pi]$ denotes the angular interval of the cut.

Everywhere except the cut the wave function obeys a free Schr\"{o}dinger equation 
\begin{equation}
\left (\Delta +k^2 \right )\Psi=0 .
\label{schrodinger}
\end{equation}  
Inside the cut circle the wave function can be expanded into series of regular elementary solutions of (\ref{schrodinger}). We find it convenient to use the Bessel function basis
\begin{equation}
\Psi^{(-)}(kr,\phi)=\sum_{m=-\infty}^{\infty}c_m J_{|m|}(kr)\mathrm{e}^{\mathrm{i}m\phi} .
\label{besselbasis}
\end{equation}
Let us denote the wave function outside the circle by $\Psi^{(+)}(kr,\phi)$.
To implement the boundary conditions  (\ref{AB_function}) and (\ref{AB_derivative}), one multiplies both equations by $\mathrm{e}^{\mathrm{i}m\phi}$ and integrates over $\phi$ from $0$ to $2\pi$  for all integer $m$ from $-\infty$ to $\infty$.  One concludes that
\begin{equation}
\fl
J_{|m|}^{\prime}(x)\int_0^{2\pi}\mathrm{e}^{-\mathrm{i}m\phi}\Psi^{(+)}(x,\phi)f(\phi)\mathrm{d}\phi
-J_{|m|}(x)\int_0^{2\pi}\mathrm{e}^{-\mathrm{i}m\phi} \frac{\partial}{\partial x}\Big ( \Psi^{(+)}(x,\phi)\Big )f(\phi)\mathrm{d}\phi=0
\label{eqs}
\end{equation}
where $x=kR$. 

The usual way to exploit these conditions (cf.~\cite{MMO})  is to expand   $\Psi^{(+)}(r,\phi)$  into a series of  Hankel functions provided that the incident wave is fixed:
\begin{equation}
\Psi^{(+)}_l(kr,\phi)=J_{|l|}(kr)\mathrm{e}^{\mathrm{i}l\phi}+\sum_{n=-\infty}^{\infty}t_n(l) H_{|n|}^{(1)}(kr) \mathrm{e}^{\mathrm{i}n\phi}.
\label{elementary}
\end{equation}
From (\ref{eqs}) it follows that the coefficients $t_n(l)$  obey an infinite system of equations
\begin{equation}
\sum_{n=-\infty}^{\infty}\left (J_{|m|}^{\prime}H_{|n|}^{(1)}-J_{|m|} H_{|n|}^{(1)\prime}\right )A_{mn}t_n(l)=-
(J_{|m|}^{\prime}J_{|l|}-J_{|m|} J_{|l|}^{\prime})A_{ml}\ ,
\label{systems_eqs}
\end{equation}
where the  $A_{mn}$'s are the Fourier coefficients of $f(\phi)$ in (\ref{phi})
\begin{equation}
A_{mn}=\frac{1}{2\pi}\int_0^{2\pi}f(\phi)\mathrm{e}^{\mathrm{i}(n-m)\phi}\mathrm{d}\phi=\mathrm{e}^{\mathrm{i}\pi\alpha}\left \{ \begin{array}{cr} \cos \pi \alpha &  (m=n)\\
\frac{\sin \pi \alpha}{\pi} \frac{1-(-1)^{n-m}}{n-m} & (m\neq n)\end{array}\right. .
\label{A_mn} 
\end{equation}
Here and below, when the argument of Bessel functions is not specified,
it is meant to be equal to  $kR\equiv x$. It is convenient  to rescale the $t_n(l)$'s as
\begin{equation}
t_n(l)=\frac{J_{|l|}(x)}{H_{|n|}^{(1)}(x)}y_n(l)
\label{definition}
\end{equation}
so that the $y_n(l)$'s obey the system of equations
\begin{equation}
\sum_{n=-\infty}^{\infty}\left (\frac{J_m^{\prime}}{J_m} -\frac{H_n^{(1)\prime}}{H_n^{(1)}}\right )A_{mn}y_n(l)=-
\left (\frac{J_m^{\prime}}{J_m}- \frac{J_l^{\prime}}{J_l}\right )A_{ml}\ .
\label{rescaled}
\end{equation} 

From figure~\ref{fig_cut} it is clear that the scattering problem is symmetric with respect to a reflection in the $y$-axis. Namely, if $\Psi(r,\phi)$ is a solution then $\Psi(r,\pi-\phi)$ is also a solution. This symmetry is a consequence of the invariance of the function $f(\phi)$ under the transformation
\begin{equation}
f(\pi-\phi)=f(\phi).
\label{invariance}
\end{equation}
It manifests itself in a symmetry of the matrix elements $A_{mn}$
\begin{equation}
A_{-m,-n}=(-1)^{n-m}A_{mn}
\end{equation}
and of the expansion coefficients $y_n(l)$
\begin{equation}
y_{-n}(-l)=(-1)^{n+l} y_n(l) .
\label{zero_symmetry}
\end{equation}
The matrix $A_{mn}$ has two more symmetries. A trivial one corresponds to an integer shift $\alpha\to 1+\alpha$, which leaves $A_{mn}$ and $y_n(l)$ invariant:
\begin{equation}
y_n(l;\alpha)=y_n(l;1+\alpha).
\end{equation}
This invariance reflects the well known fact that the Aharonov-Bohm effect depends only on the fractional part of $\alpha$. 

The second symmetry is related with the transformation $\alpha\to -\alpha$. From (\ref{A_mn}) one concludes that
\begin{equation}
A_{m n}(-\alpha)=\mathrm{e}^{-2\pi \mathrm{i}\alpha}A_{-m-n}(\alpha)=(-1)^{m-n}\mathrm{e}^{-2\pi \mathrm{i}\alpha} A_{m n}(\alpha)
\end{equation}
which  leads to the symmetry for the  coefficients $y_n(l)$
\begin{equation} 
y_n(l;-\alpha)=y_{-n}(-l;\alpha)=(-1)^{n+l}y_n(l;\alpha).
\label{second_symmetry}
\end{equation}
The knowledge of coefficients $t_n(l)$ (or $y_n(l)$) determines all other quantities.  Of particular interest is the amplitude of scattering of a plane wave on the  two vortices. To determine it, one has to find   a solution which at large distances from the vortices is the superposition of an incoming  plane wave  in the direction $\theta$ and a circular symmetric outgoing wave 
\begin{equation}
\Psi^{(+)}(kr,\phi)\underset{r\to \infty}{\longrightarrow}\mathrm{e}^{\mathrm{i}kr\cos (\theta-\phi)}+\sqrt{\frac{2}{\pi k r}}\mathrm{e}^{\mathrm{i}kr-\mathrm{i}\pi/4}F(\theta,\phi) \ .
\label{asymptotic}
\end{equation} 
Expanding the incoming plane wave into a series of Bessel functions (see, e.g., \cite{bateman})
\begin{equation}\label{expansion_wave_minus}
\mathrm{e}^{\mathrm{i}kr\cos (\theta-\phi)}=\sum_{l=-\infty}^{\infty}\mathrm{i}^{|l|} 
J_{|l|}(kr)\mathrm{e}^{\mathrm{i}l(\phi-\theta)}
\end{equation}
 the scattering wave function can be expressed through   $\Psi^{(+)}_l(kr,\phi)$ defined in (\ref{elementary}) as
\begin{equation}
\Psi^{(+)}(kr,\phi)=\sum_{l=-\infty}^{\infty}\mathrm{i}^{|l|} \Psi^{(+)}_l(kr,\phi)\mathrm{e}^{-\mathrm{i}l\theta} =\mathrm{e}^{\mathrm{i}kr\cos (\theta-\phi)}+ \Psi^{(\mathrm{ref})}(kr,\phi)\ . 
\label{field}
\end{equation}
 $\Psi^{(\mathrm{ref})}(kr,\phi)$ is the reflected field given by a series of  Hankel functions of the first kind
\begin{equation}
\Psi^{(\mathrm{ref})}(kr,\phi)= \sum_{n,l=-\infty}^{\infty}\mathrm{i}^{|l|} t_n(l)  H_{|n|}^{(1)}(kr) \mathrm{e}^{\mathrm{i}n\phi-\mathrm{i}l\theta} \ . 
\label{Hankel_series}
\end{equation}
From the asymptotic behaviour of the Hankel functions (see, e.g., \cite{bateman})
\begin{equation}
H_n^{(1)}(kr)\underset{r\to \infty}{\longrightarrow} \sqrt{\frac{2}{\pi k r}}
\mathrm{e}^{\mathrm{i}\left (kr -\pi n/2-\pi/4 \right )}
\end{equation}
one gets that the scattering amplitude $F(\theta,\phi)$ in (\ref{asymptotic}) is 
\begin{equation}
F(\theta,\phi) =\sum_{n,l=-\infty}^{\infty}\mathrm{i}^{|l|-|n|} t_n(l)\mathrm{e}^{\mathrm{i}n \phi-\mathrm{i}l\theta}\ .
\label{amplitude}
\end{equation}
The symmetries (\ref{zero_symmetry}) and (\ref{second_symmetry}) lead to the symmetries of the scattering amplitude
\begin{equation}
F(\pi-\theta,\pi-\phi)=F(\theta,\phi)
\label{sym_1}
\end{equation}
and 
\begin{equation}
F(\theta,\phi;-\alpha)=F(\theta,\phi;1-\alpha)=F(\pi+\theta,\pi+\phi;\alpha)\ .
\label{sym_2}
\end{equation}
These relations make it possible to restrict the value of flux to
$0<\alpha\leq \frac{1}{2}$.
 
\section{Numerical results}\label{nume}

We proceed to calculate the scattering amplitude numerically,
by solving the systems (\ref{rescaled}) for $y_n(l)$, for different values of $l$,
and substituting the result into (\ref{definition}) and then (\ref{amplitude}). From physical considerations it follows that contributions of partial waves with $|l| \ll kR$ have to be small.
In the calculations below we restrict ourselves to $|l| \le 10kR$ and check that
higher waves do not change the results noticeably.  

Truncation in $n$ is more complicated. In order to find the coefficients themselves with good precision,
one has to retain much higher values of $n$ in the systems (\ref{rescaled}), 
$|n| \le N$ with $N \sim 100$.
Even then, the precision attained at technically feasible values of $N$ is not sufficient, and one has to extrapolate the results to $N \to \infty$.

For a given finite $N$, we truncate the infinite sum in (\ref{rescaled})
to $n = -N+1, \ldots, N$, so that the system, for a given $l$, consists of
$2N$ equations for as many variables.

It turns out that convergence in $N$ is improved  by employing the technique used in \cite{MMO}, discretizing the boundary conditions. The issue at hand is that by choosing $2N$ coefficients $t_n(l)$ in
(\ref{elementary}) and, respectively, $2N$ coefficients $c_m$
in (\ref{besselbasis}), one can, in general, satisfy the boundary conditions
(\ref{AB_function})--(\ref{AB_derivative}) exactly for no more than $2N$ discrete
values of $\phi$, rather than for any $\phi$. A convenient recourse is to enforce those conditions at
$\phi_k = (k-\frac{1}{2})\frac{\pi}{N}$, $k=1,\ldots,2N$
(these $2N$ points are distributed uniformly on the circle and avoid as much as possible
the locations of the fluxes, where the wave function is singular).
Then, integration over $\phi$ in (\ref{eqs}) has to be replaced with summation over $\phi_k$.
It is easy to see that the only change this entails is a modified expression
for $A_{mn}$, namely
\begin{equation}
A_{mn}=\mathrm{e}^{\mathrm{i}\pi\alpha}\left \{ \begin{array}{cr} \cos \pi \alpha &  (m=n)\\
\frac{\sin \pi \alpha}{2N} \frac{1-(-1)^{n-m}}{\sin \frac{(n-m)\pi}{2N}} & (m\neq n)\end{array}\right. .
\label{A_mn_discrete}
\end{equation}
For $N \to \infty$, one recovers (\ref{A_mn}), whereas for a given finite $N$,
the resulting amplitude is closer to the infinite-$N$ one than the
one obtained without this modification.

Convergence in $N$ of the resulting scattering amplitude is slow,
due to the usual problem inherent in Aharonov-Bohm or anyon numerics:
We are trying to represent a wave function which has a fractional power behaviour
at the positions of the vortices (cf.~(\ref{singular_limits})),
as an expansion in terms of regular wave functions.
The corresponding coefficients fall off slowly.
We have found, however, that for all values of $kR$, $\theta$, and $\alpha$
considered, and for any $\phi \in [-\pi, \pi]$, the $N$ dependence of the amplitude
fits an empirical formula
\begin{equation}
F(\theta, \phi; N) \simeq F(\theta, \phi) + \frac{a_1}{N^{\alpha}}
+ \frac{a_2}{N^{1-\alpha}} + \frac{a_3}{N}+\frac{a_4}{N^{1+\alpha}}
\label{general_fit}
\end{equation}
well, even for rather small $N$. (We have used the values $N = 40 \ldots 100$ for the fitting.)

\begin{figure}
\begin{center}
\includegraphics[angle=-90,width=.99\linewidth]{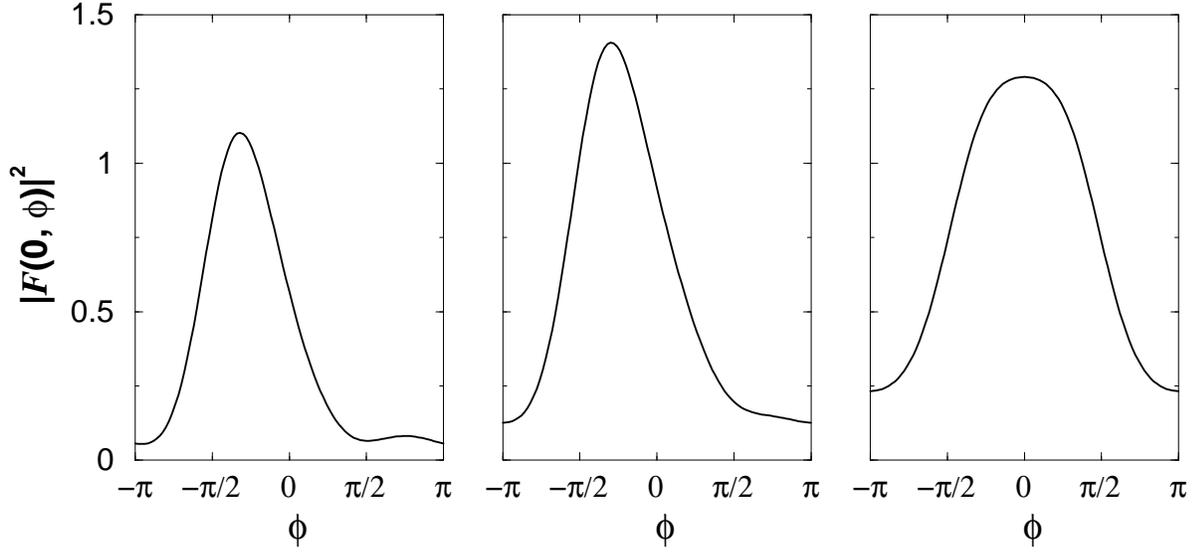}
\end{center}
\caption{Modulus squared of the scattering amplitude for $kR=1$ and the incident angle $\theta=0$. Left: $\alpha=1/4$, center: $\alpha=1/3$, right: $\alpha=1/2$. }
\label{fig_theta_0}
\end{figure}

\begin{figure}
\begin{center}
\includegraphics[angle=-90,width=.99\linewidth]{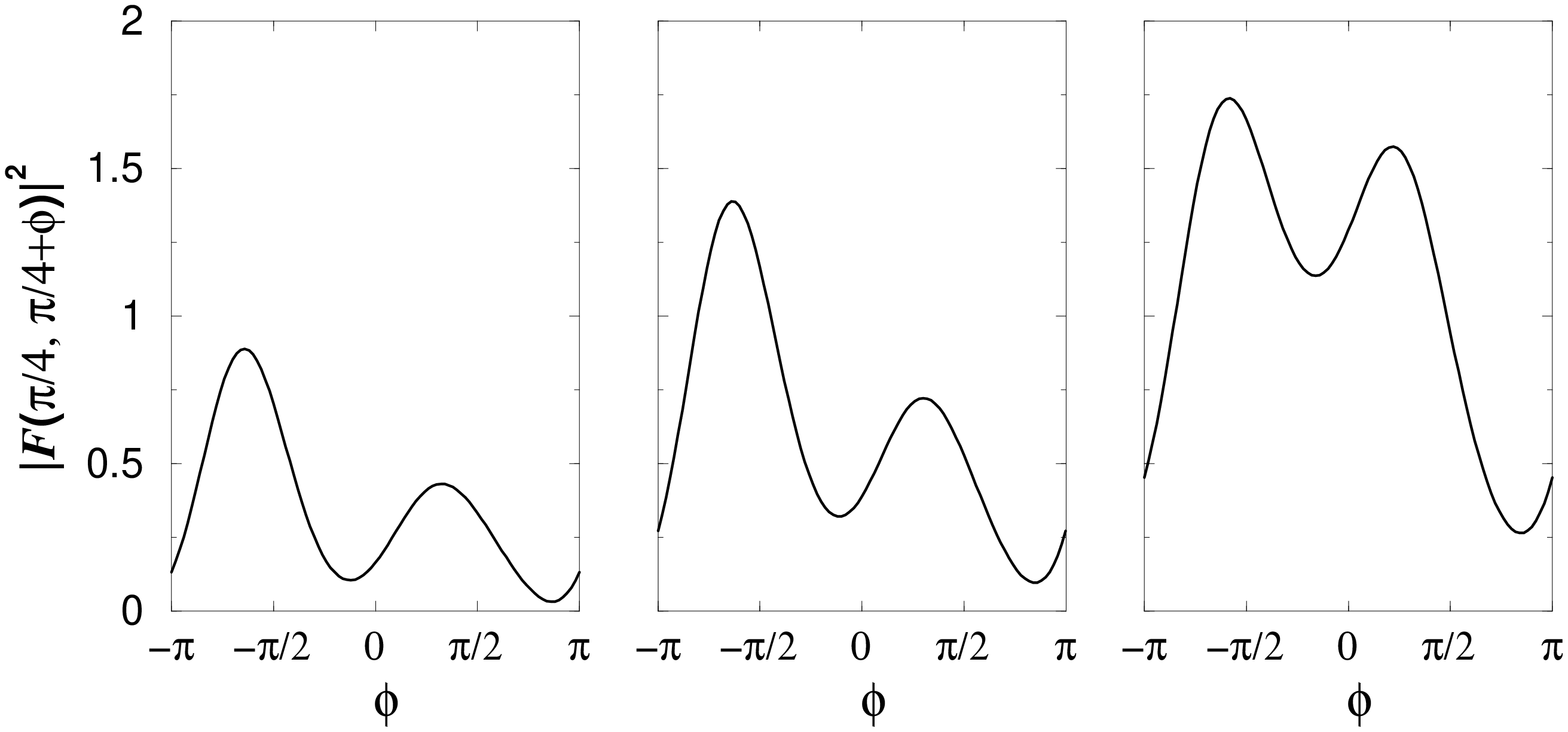}
\end{center}
\caption{The same as in figure~\ref{fig_theta_0} but for the incident angle $\theta=\pi/4$. }
\label{fig_theta_pi_4}
\end{figure}

\begin{figure}
\begin{center}
\includegraphics[angle=-90,width=.99\linewidth]{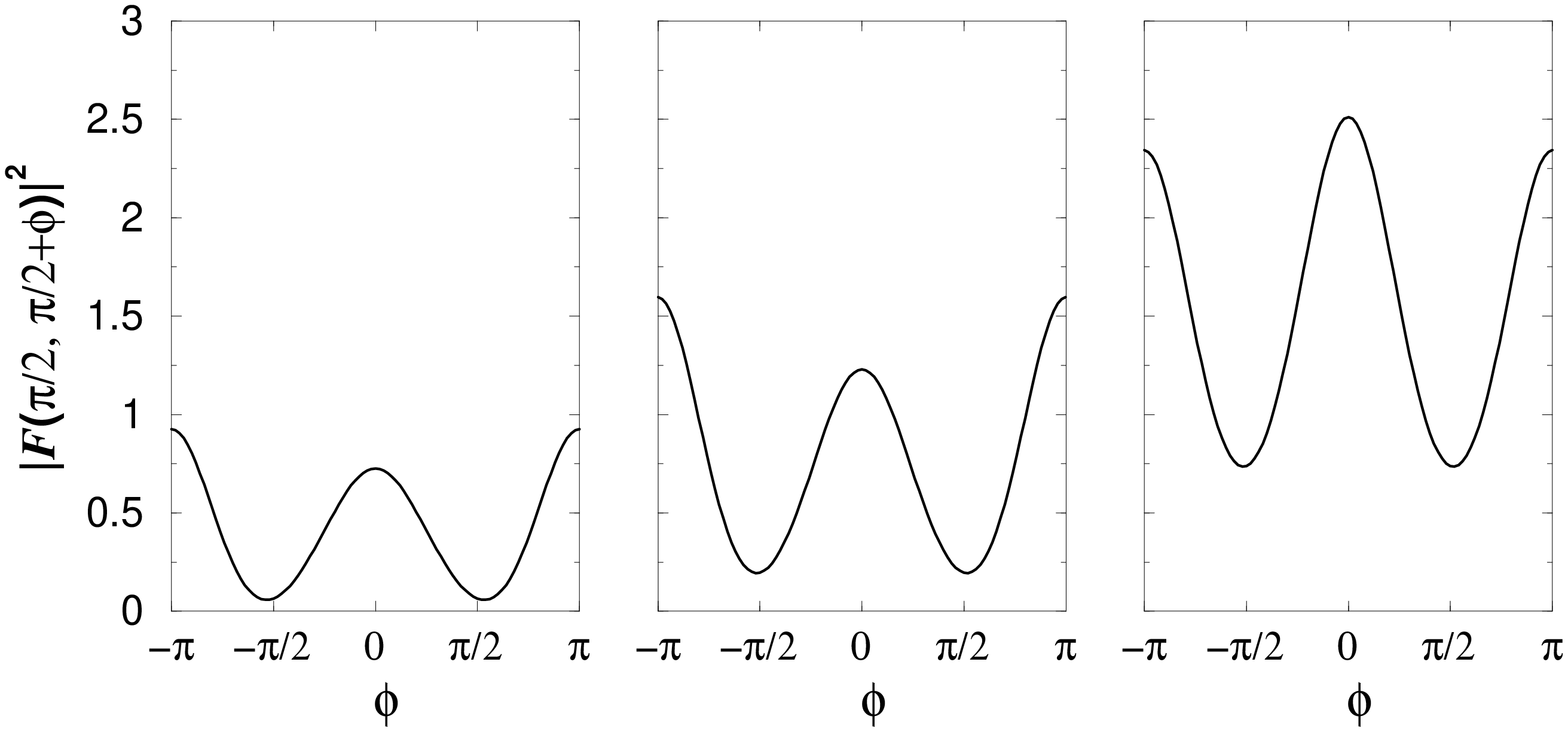}
\end{center}
\caption{The same as in figure~\ref{fig_theta_0} but for but for the incident angle  $\theta=\pi/2$. }
\label{fig_theta_pi_2}
\end{figure}

In figures~\ref{fig_theta_0}-\ref{fig_theta_pi_2} we present the results of numerical calculations for the amplitude   of the scattering on two Aharonov-Bohm vortices for fluxes $\alpha=1/4$, $1/3$, and $1/2$ with $kR=1$ and with incident angles $\theta=0$, $\pi/4$, $\pi/2$. For clarity, in these figures we count the reflection angle
from the direction of the incident wave, i.e., we plot $F(\theta,\theta+\phi)$ instead of  $F(\theta,\phi)$.

A notable feature of the results is the absence of symmetry
with respect to the reflection $\phi \to -\phi$.
This is in contrast to the standard Aharonov-Bohm scattering, where, despite the fact that
chiral symmetry is broken by the presence of the vortex, the amplitude remains symmetric
under such reflection. One could have expected the same here for $\theta = 0$,
when the geometry of the system is just as symmetric with respect to the oncoming wave
as it is with a single vortex.
However, in general this turns out not to be the case because for the invariance the spacial reflection has to be accompanied by interchanging of the vortices. For $\theta=0$ only the case $\alpha=1/2$ is symmetric with respect to inversion $\phi\to -\phi$ (cf. (\ref{sym_1}) and (\ref{sym_2})). In general, with positive $\alpha$ the reflected wave is more likely to deviate to the right (negative $\phi$).
Only for $\theta = \pi/2$ is the amplitude symmetric, as anticipated (cf.~(\ref{sym_1})).

\section{Small $\alpha$ limit}\label{small_alpha}

To find the leading term of the scattering amplitude in the   limit  $\alpha\to 0$, it is convenient first to transform  the boundary conditions to another form which is of independent interest. Multiplying both sides of (\ref{eqs})  by 
$\mathrm{i}^{|m|}\mathrm{e}^{\mathrm{i}m (\theta^{\prime}+\pi)}$, summing over $m$,  and using again the identity (\ref{expansion_wave_minus}), one concludes that $ \Psi^{(+)}$ has  to obey 
\begin{equation}
\fl
\int_0^{2\pi} \left [\mathrm{e}^{-\mathrm{i}x\cos (\theta^{\prime}-\phi)} \frac{\partial}{\partial x}\Big ( \Psi^{(+)}(x,\phi)\Big )  - \Psi^{(+)}(x,\phi)  \frac{\partial}{\partial x}\Big (\mathrm{e}^{-\mathrm{i}x\cos (\theta^{\prime}-\phi)}\Big )  \right ]f(\phi)\mathrm{d}\phi=0\ .
\label{green}
\end{equation}
This equation can also be obtained by a direct application of the Green identity inside the circle. 

For the pure scattering setup one has to take $\Psi^{(+)}(kr, \phi)$ as in (\ref{field}).  
From (\ref{green}) it follows that  the reflected field obeys the equation
\begin{eqnarray} 
\fl \int_0^{2\pi}\left [\mathrm{e}^{-\mathrm{i}x\cos (\theta^{\prime}-\phi)} \frac{\partial}{\partial x}\Big (\Psi^{(\mathrm{ref})}(x,\phi)\Big )\right .   &-&\left .\Psi^{(\mathrm{ref})}(x,\phi) \frac{\partial}{\partial x}\Big (\mathrm{e}^{-\mathrm{i}x\cos (\theta^{\prime}-\phi)}\Big )   \right ]f(\phi) \mathrm{d}\phi\nonumber\\
&=&J(\theta,\theta^{\prime})
\label{integral}
\end{eqnarray}
where 
\begin{equation}
\fl
J(\theta,\theta^{\prime})=\int_0^{2\pi} \left [\mathrm{e}^{-\mathrm{i}x\cos (\theta^{\prime}-\phi)} \frac{\partial}{\partial x}\left ( \mathrm{e}^{\mathrm{i}x\cos (\theta-\phi)} \right )  - \frac{\partial}{\partial x}\left (\mathrm{e}^{-\mathrm{i}x\cos (\theta^{\prime}-\phi)}\right )\mathrm{e}^{\mathrm{i}x\cos (\theta-\phi)}  \right ]f(\phi) \mathrm{d}\phi\, .
\label{exact}
\end{equation}
The expression in the square brackets is a total derivative:
\begin{eqnarray}
& &\mathrm{e}^{-\mathrm{i}x \cos (\theta^{\prime}-\phi)} \frac{\partial}{\partial x}
\left ( \mathrm{e}^{\mathrm{i}x\cos (\theta-\phi)} \right )  
-\frac{\partial}{\partial x}\left ( \mathrm{e}^{-\mathrm{i}x\cos (\theta^{\prime}-\phi)}\right ) \mathrm{e}^{\mathrm{i}x\cos (\theta-\phi)} 
\nonumber\\ 
&=&\frac{2\mathrm{i}}{x}\cot \left (\frac{\theta-\theta^{\prime}}{2}\right ) \frac{\partial}{\partial \phi}
\exp \left [ -2 \mathrm{i}x \sin \left (\frac{\theta+\theta^{\prime}}{2}-\phi \right )  \sin \left (\frac{\theta-\theta^{\prime}}{2}\right )\right ]
\end{eqnarray}
so the integration  in (\ref{integral})  for $f(\phi)$ given in (\ref{phi}) easily follows as
\begin{equation}
J(\theta,\theta^{\prime})=\frac{4}{x}\mathrm{e}^{\mathrm{i}\pi\alpha}\sin \pi \alpha \cot \frac{\theta-\theta^{\prime}}{2}\sin \left (2x\sin \frac{\theta-\theta^{\prime}}{2}\sin \frac{\theta+\theta^{\prime}}{2}  \right )\ .
\label{J}
\end{equation}
Using the second expression for $f(\phi)$ in (\ref{phi}), (\ref{integral}) can be rewritten as
\begin{eqnarray}
\Phi(\theta,\theta^{\prime})&=&-2\mathrm{i}\mathrm{e}^{\pi \mathrm{i} \alpha}\sin \pi \alpha \int_{\pi}^{2\pi} \left [\mathrm{e}^{-\mathrm{i}x\cos (\theta^{\prime}-\phi)} \frac{\partial}{\partial x} \Psi^{(\mathrm{ref})}(r,\phi) \right .\nonumber\\
 &-& \left .\frac{\partial}{\partial x}\left (\mathrm{e}^{-\mathrm{i}x\cos (\theta^{\prime}-\phi)}\right ) \Psi^{(\mathrm{ref})}(r,\phi)  \right ]
\mathrm{d}\phi\nonumber\\
 &+&
\frac{4}{x}\mathrm{e}^{\mathrm{i}\pi\alpha}\sin \pi \alpha \cot \frac{\theta-\theta^{\prime}}{2}\sin \left (2x\sin \frac{\theta-\theta^{\prime}}{2}\sin \frac{\theta+\theta^{\prime}}{2}  \right )
\label{small_alpha_series}
\end{eqnarray}
where 
\begin{equation}
\fl 
\Phi(\theta,\theta^{\prime})=\int_{0}^{2\pi} \left [\mathrm{e}^{-\mathrm{i}x\cos (\theta^{\prime}-\phi)} \frac{\partial}{\partial x} \Psi^{(\mathrm{ref})}(r,\phi) - \frac{\partial}{\partial x}\left (\mathrm{e}^{-\mathrm{i}x\cos (\theta^{\prime}-\phi)}\right ) \Psi^{(\mathrm{ref})}(r,\phi)  \right ]
\mathrm{d}\phi\ .
\label{F_theta}
\end{equation}
It is well known (and easily  checked either by the Green representation of the reflected field or by the direct substitution of (\ref{Hankel_series}) in (\ref{F_theta}))  that $\Phi(\theta,\theta^{\prime})$ is proportional to  the scattering amplitude for the reflected field
\begin{equation}
\Phi(\theta,\theta^{\prime})=\frac{4\mathrm{i}}{x}F(\theta,\theta^{\prime})\ .
\end{equation}
If one assumes that at small $\alpha$ (hence, small $\sin \pi \alpha$) the wave function $\Psi^{(\mathrm{ref})}(r,\phi)$ can be expanded into a series in $\alpha$ 
as
\begin{equation}
\Psi^{(\mathrm{ref})}(r,\phi) =\alpha \Psi_1(r,\phi)+\alpha^2 \Psi_2(r,\phi) +\ldots 
\end{equation}
(the zero order term is absent since when $\alpha$ is an integer there is no Aharonov-Bohm scattering)
and takes into account that the right-hand side of   (\ref{small_alpha_series}) is proportional to $\sin \pi \alpha$, then  the scattering amplitude in the leading order can be approximated 
by
\begin{equation}
F(\theta,\theta^{\prime})\approx 
-\mathrm{i}\sin \pi \alpha \cot \frac{\theta-\theta^{\prime}}{2}\sin \left (2x\sin \frac{\theta-\theta^{\prime}}{2}\sin \frac{\theta+\theta^{\prime}}{2}  \right )\ .
\label{small}
\end{equation}
The existence of a term proportional to $N^{-\alpha}$ in the extrapolation formula (\ref{general_fit}) renders numerical calculations as very small $\alpha$ with  $N$ of the order of a few hundred uncertain. Fortunately, the analytical formula (\ref{small}) for the dominant contribution at small $\alpha$  (\ref{small}) is regular and does not contain singularities at the vortex positions. This means that the coefficient in front of $N^{-\alpha}$ is, at least, of the second order in $\alpha$. Therefore, at very small $\alpha$ we 
drop that term from the fit (\ref{general_fit}).

\begin{figure}
\begin{center}
\includegraphics[angle=-90, width=.99\linewidth]{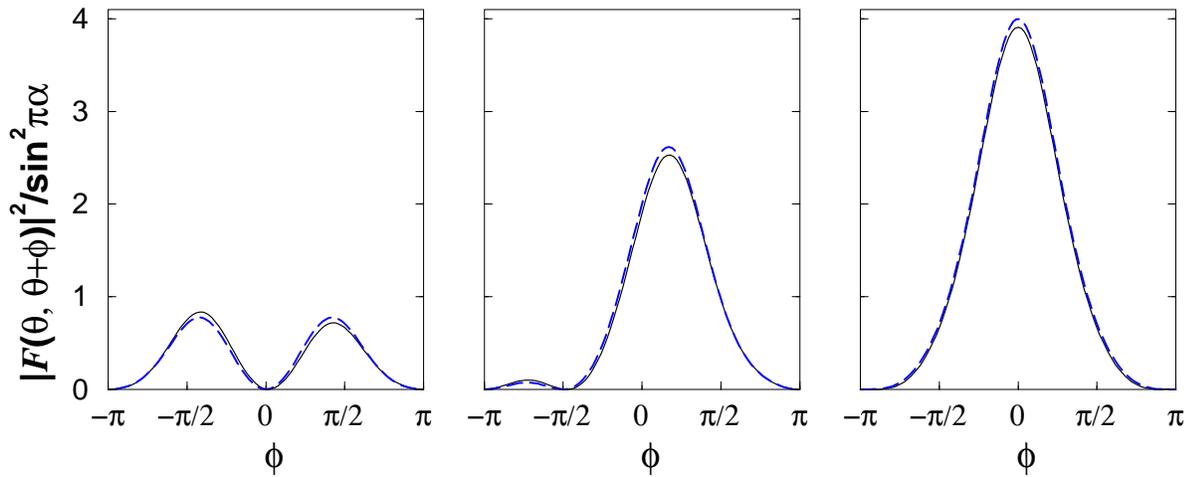}
\end{center}
\caption{Scattering amplitudes on two vortices with $\alpha=.01$ and $kR=1$ for different incident angle $\theta$ versus the reflection angle $\phi$. Left:  $\theta=0$, center: $\theta=\pi/4$, right: $\theta=\pi/2$. Solid lines are  direct numerical solutions of (\ref{rescaled}). Dashed lines are calculated from the analytic formula~(\ref{small}).}
\label{fig_1}
\end{figure}

The results obtained in this manner are  presented  in figure~\ref{fig_1} for different incident angles and compared  with (\ref{small}) for small $\alpha$. For clarity, we divide the  scattering amplitude by $\sin \pi \alpha$ and, like in the previous Section, count the reflection angle from the direction of the incident wave.  The agreement is quite good and the differences between  both curves are of the order of the terms which have been neglected (i.e., $\sin \pi \alpha$).   

\section{Small $kR$ limit}\label{small_kR_case}
Another interesting limit corresponds to the vortices  close to each other, so that $kR\ll 1$. 
In this case it is convenient to use the rescaled  variables (\ref{rescaled}) and to expand all quantities at small $x$. One gets  (see, e.g., \cite{bateman})
\begin{equation}
\frac{J_m^{\prime}(x)}{J_m(x)}\underset{x\to 0}{\longrightarrow}\frac{|m|}{x}\ ,\;\; \;
\frac{H_n^{(1)\prime}(x)}{H_n^{(1)}(x)}\underset{x\to 0}{\longrightarrow}\frac{-|n|+\delta_{n 0}\rho}{x}
\end{equation}
with
\begin{equation}
\rho=\frac{2\mathrm{i}}{\pi+2\mathrm{i}(\ln (x/2)+\gamma)}
\label{rho}
\end{equation}
($\gamma$ is the Euler constant). 

In this limit   equations (\ref{rescaled}) take  the form
\begin{equation}
\sum_{n=-\infty}^{\infty}(|m|+|n|-\delta_{n0}\rho )A_{mn}y_n(l)=-(|m|-|l|)A_{ml}
\label{limit_system}
\end{equation}
so that the  dependence on $x=kR$ is entirely contained in $\rho$. On the other hand, from (\ref{definition}) it follows  that when $x\to 0$,  $t_n(l)\sim x^{|n|+|l|}y_n(l)$. Therefore at small $x$ the dominant contribution to the scattering amplitude  comes from  the $S$-wave amplitude  ($l=0$), more precisely from $y_0(0)$, as it is usual  for the scattering on small-size objects (see, e.g., \cite{albeverio}). Nevertheless, to determine $y_0(0)$, one still has to solve an infinite system of equations (\ref{limit_system}) with $l=0$:
\begin{equation}
\sum_{n=-\infty}^{\infty}(|m|+|n|-\delta_{n0}\rho )A_{mn}y_n(0)=-|m|A_{m0}\ .
\label{main}
\end{equation}
The scattering amplitude (\ref{amplitude}) is dominated at small $x$ by $y_0(0)$ 
\begin{equation}
F(\theta,\phi)\underset{x\to 0}{\longrightarrow} t_0(0)=\frac{y_0(0)}{H_0^{(1)}(x)}\approx 
-\mathrm{i}\frac{\pi}{2}\rho y_0(0)\ 
\label{limit_F}
\end{equation}
where we  used that, at small $x$,  $H_0^{(1)}(x)\approx 2\mathrm{i}/(\pi \rho)$  with $\rho$ given by (\ref{rho}) (see, e.g., \cite{bateman}). 
 
From (\ref{zero_symmetry}) it follows that 
\begin{equation}
y_{-n}(0)=(-1)^n y_n(0)
\label{symmetry_y}
\end{equation}
and one can transform (\ref{main}) to a system of  equations involving the $y_n(0)$'s with non-negative $n$  only. 
When $m=0$ one obtains
\begin{equation}
\cos \pi \alpha \ \rho y_0-\frac{2\sin \pi \alpha}{\pi}\sum_{n=1}^{\infty}(1-(-1)^n)y_n(0)=0\ .
\label{first}
\end{equation}
For positive  $m$, straightforward calculations demonstrate that due to symmetry (\ref{symmetry_y})   
\begin{eqnarray}
&& \sum_{\substack{n=-\infty, \; n\neq 0} }^{\infty}(m+|n|)A_{mn}y_n(0) \nonumber\\
&&=2m\sum_{n=1}^{\infty}A_{mn}y_n(0) + \frac{\sin \pi \alpha}{\pi }(1+(-1)^{m})\sum_{n=1}^{\infty}(1-(-1)^n)y_n(0)\ .
\end{eqnarray}
Using (\ref{first}) and adding the $n=0$ terms,  one gets that (\ref{main}) with positive $m$ is equivalent to
\begin{eqnarray}
\sum_{n=1}^{\infty}A_{mn}y_n(0)&=&\frac{\sin \pi \alpha (1-(-1)^{m})}{2\pi m}(1+y_0(0))\nonumber\\
& -&
\left (\frac{\sin \pi \alpha (1-(-1)^{m})}{2\pi m^2}+
\frac{\cos \pi \alpha ( 1+(-1)^{m})}{4m}\right )\rho y_0(0) .
\label{second}
\end{eqnarray}
All the dependence on $x=kR$ in this equation is in $\rho$  only.
 
It is convenient to redefine the variables $y_n(0)$ with $n\geq 1$ as   
\begin{equation}
y_n(0)=\frac{1}{2} x_n(1+y_0(0))
\label{y_n}
\end{equation}
and to set
\begin{equation}
x_0=1 
\label{norm}
\end{equation}
so that (\ref{second}) takes the form
\begin{equation}
\sum_{n=0}^{\infty}A_{mn}x_n=\xi f_m 
\label{HR}
\end{equation}
where 
\begin{equation}
\xi=\frac{2\rho y_0(0)}{1+y_0(0)}
\end{equation}
and 
\begin{equation}
f_m=-\frac{\sin \pi \alpha (1-(-1)^{m})}{2\pi m^2}-\frac{\cos \pi \alpha ( 1+(-1)^{m})}{4m}\ .
\label{f_m}
\end{equation}
Similarly  (\ref{first}) becomes 
\begin{equation}
\frac{\pi}{2} \xi \cot \pi \alpha=\sum_{n=1}^{\infty}x_n(1-(-1)^n)\ . 
\label{first_HR}
\end{equation}  
The system  (\ref{HR}) is linear,  so its solution  has  the form 
\begin{equation}
x_n=a_n+b_n \xi
\end{equation}
where $a_n$ and $b_n$ depend  on $\alpha$ but not  on $\rho$ and not on $y_0(0)$.  From  (\ref{first_HR}) one concludes that $\xi$ has to be determined by
\begin{equation}
\frac{\pi}{2} \xi \cot \pi \alpha=\sum_{n=1}^{\infty}a_n(1-(-1)^n)+\xi \sum_{n=1}^{\infty}b_n(1-(-1)^n\ . 
\label{xi_A_B}
\end{equation}
Solving (\ref{xi_A_B}) for $\xi$ one finds
\begin{equation}
\frac{\rho y_0(0)}{1+y_0(0)}=-\frac{1}{\beta(\alpha)}
\label{beta}
\end{equation}
where $\beta(\alpha)$ is  real. Therefore 
\begin{equation}
y_0(0)=-\frac{1}{1+\beta(\alpha)\rho}\ .
\label{beta_alpha}
\end{equation}
where  the dependence on $kR$ and $\alpha$ has separated. 

From (\ref{limit_F}) the scattering amplitude in the small $kR$ limit is
\begin{equation}
F(\theta,\phi)\underset{kR\to 0}{\longrightarrow} -\frac{\pi}{\pi+2\mathrm{i}(\ln(kR/2)+\gamma+\beta(\alpha))}\ .
\label{small_kR}
\end{equation}
It has the standard form of a short-range scattering amplitude (e.g., the scattering on a small disk of radius $R$ with Dirichlet boundary conditions corresponding to $\beta(\alpha)=0$ (see, e.g., \cite{albeverio})). 

\begin{figure}
\begin{center}
\includegraphics[angle=-90, width=.99\linewidth]{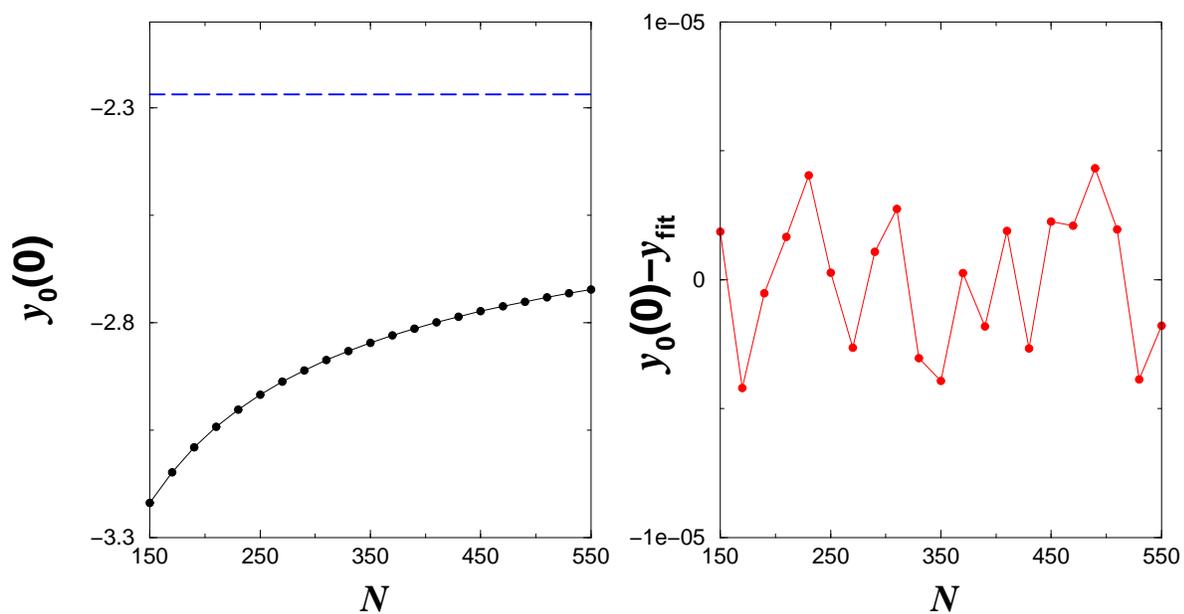}
\end{center}
\caption{On the left: red circles are the values of $y_0(0)$  calculated numerically from (\ref{main}) with $\alpha=.2$ and $\rho=.3$; the dashed line represents the limiting value of $y_0(0)$ (i.e. $a_0$ in the fit (\ref{fit})). On the right: the difference between $y_0(0)$ computed numerically at finite $N$ as on the left figure and the fit~(\ref{fit}) with fitting coefficients (\ref{data_fit}). The solid red line serves only to join the points.}
\label{fig2}
\end{figure}

A simple way of determining $\beta(\alpha)$ numerically is  to  fix  $\rho$ to a given value $\rho_0$ (say,
$\rho_0=0.3$) and to approximate the infinite system (\ref{limit_system}) by a truncated  one
\begin{equation}\label{finite_N}
\sum_{n=-N}^{N}(|m|+|n|-\delta_{n0}\rho )A_{mn}y_n(0)=-|m|A_{m0}
\end{equation}
with $N$ finite ($m$ is also taken from $-N$ to $N$). (\ref{finite_N})  is solved numerically and yields $y_0(0)$  for a given $N$ as a function of $\alpha$, from which $\beta(\alpha)$  follows  using (\ref{beta_alpha}). 
The true $\beta(\alpha)$ is then obtained by taking the limit $N\to \infty$. 
As the resulting function should have  power singularities at the vortex locations (cf.~(\ref{singular_limits}) below), the convergence with increasing $N$ is slow, all the more so when $\alpha$ is close to $0$. We found that at large $N$ a good fit of the numerical data involves the same terms as in (\ref{general_fit}) 
\begin{equation}
y_\mathrm{fit}\simeq a_0+\frac{a_1}{N^{\alpha}}+\frac{a_2}{N^{1-\alpha}}+\frac{a_3}{N}+\frac{a_4}{N^{1+\alpha}}\ .
\label{fit}
\end{equation} 
For illustration, in figure~\ref{fig2} we present on the left numerical calculations for $\alpha=.2$ and $\rho=.3$ for $N$ from $150$ to $550$. The best fit to the data corresponds to (\ref{fit}) with
\begin{equation} 
a_0\simeq -2.2688,\; a_1\simeq -.4504,\; a_2\simeq -133.8,\;
a_3\simeq 449.3,\; a_4\simeq -552.0\ .
\label{data_fit}
\end{equation}
On the right of figure~\ref{fig2}, the difference between the data and the best fit is presented. It is small ($\sim 10^{-6}$),  structureless, and can be attributed to random numerical errors,  a strong argument for the validity of the fit.  Of course, this type of extrapolation is not stable, depends on the form of the extrapolating curve,  and the results should be taken with some care.  

\begin{figure}
\begin{center}
\includegraphics[width=.5\linewidth, angle=-90]{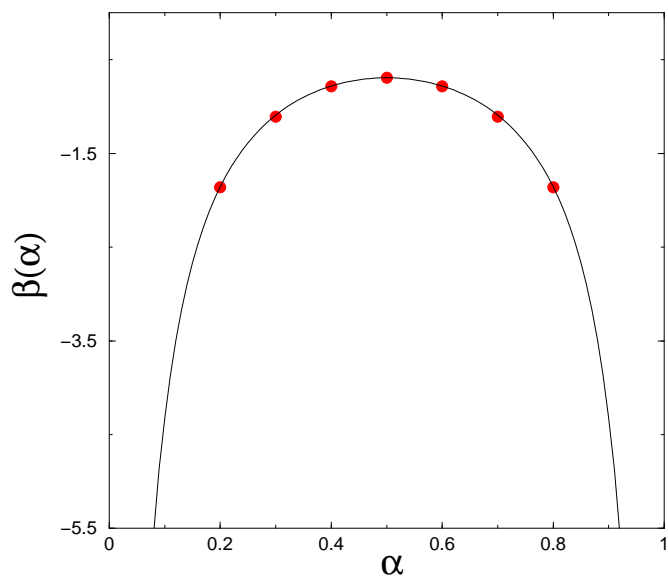}
\end{center}
\caption{Red circles: $\beta(\alpha)$ calculated numerically; black  solid line: exact formula (\ref{beta_exact}). }
\label{fig_alpha}
\end{figure}

In figure~\ref{fig_alpha} we present  numerical estimates of  $\beta(\alpha)$ for a few values of $\alpha$ calculated by the above method together with  the exact expression  (\ref{beta_exact}) which will be derived  in the next Sections.

\begin{figure}
\begin{minipage}{.48\linewidth}
\begin{center}
\includegraphics[ angle=-90, width=.99\linewidth,]{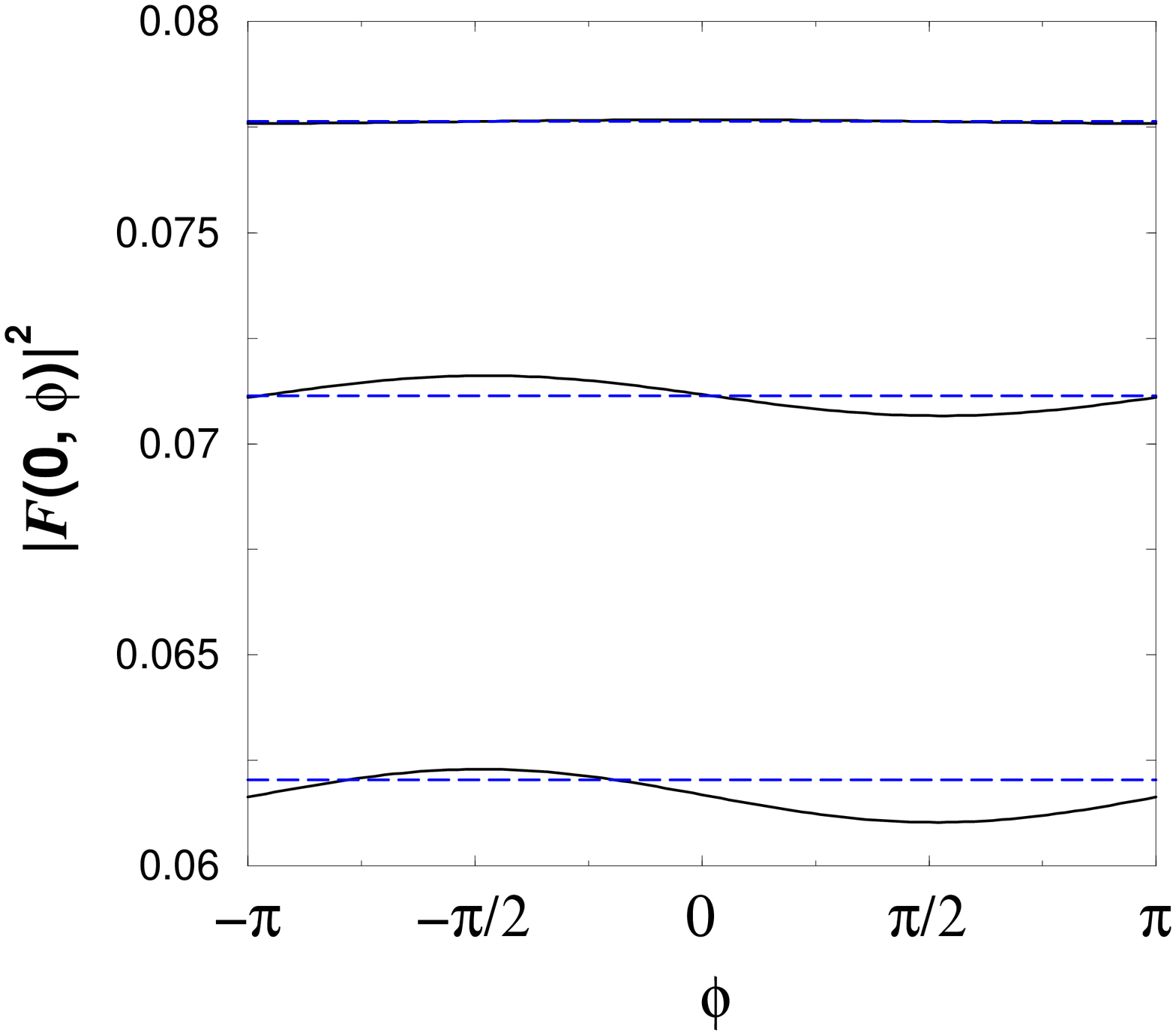}
\end{center}
\end{minipage}\hfill
\begin{minipage}{.48\linewidth}
\begin{center}
\includegraphics[ angle=-90, width=.99\linewidth,]{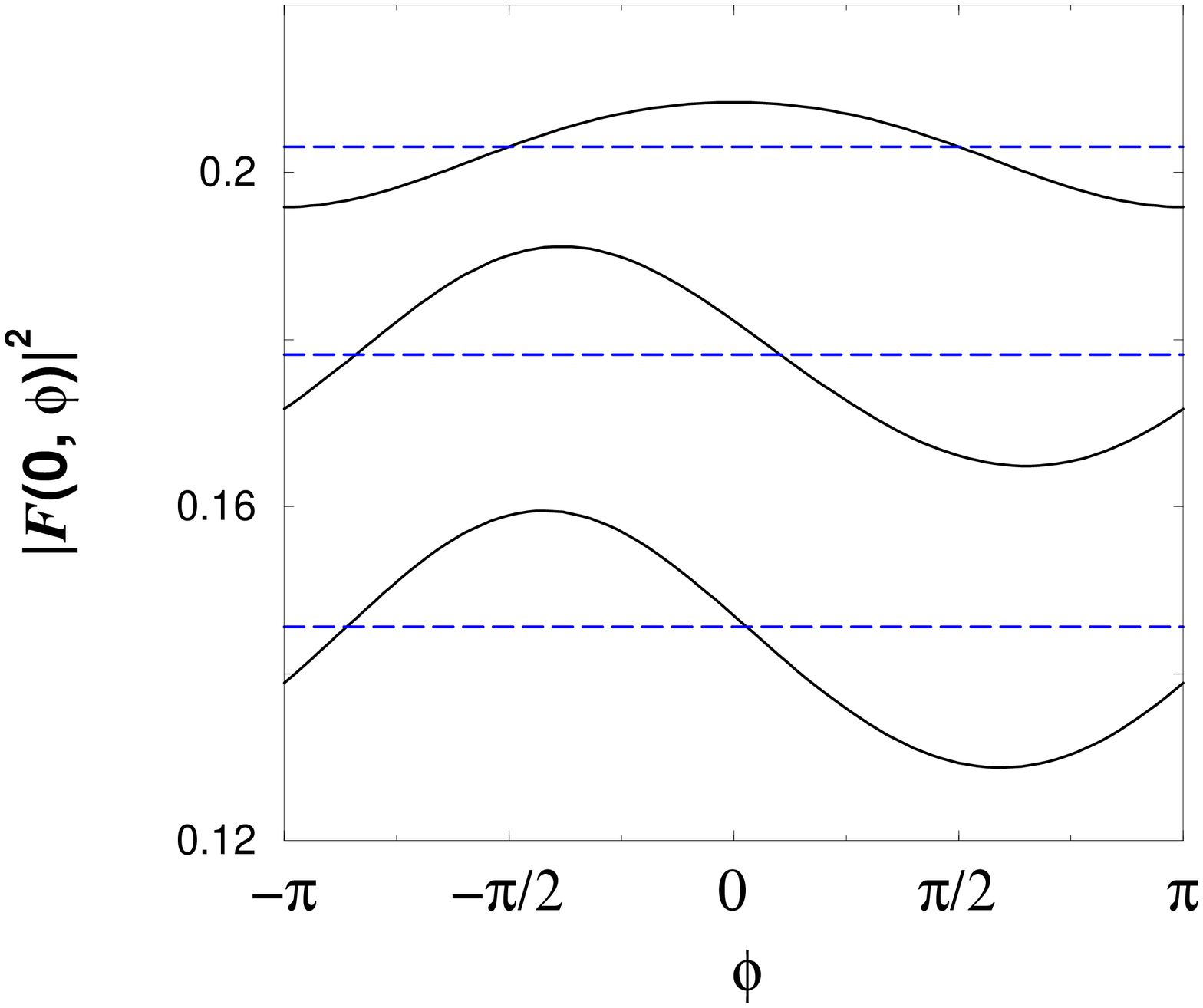}
\end{center}
\end{minipage}
\caption{Black solid lines: modulus squared of the scattering amplitude with the incident angle $\theta=0$ and different fluxes. From bottom to top: $\alpha=1/4$, $\alpha=1/3$, and $\alpha=1/2$.  Blue dashed lines are the predictions (\ref{small_kR}) for the corresponding values of $\alpha$. Left: $kR=0.01$, right: $kR=0.1$.}
\label{fig_small_kR}
\end{figure}

In figure~\ref{fig_small_kR} the scattering amplitudes calculated numerically directly from the main system of  equations (\ref{rescaled}) for $kR=0.01$ and $kR=0.1$ are plotted  for  $\alpha=1/4,\ 1/3,\ 1/2$ and the incident angle $\theta=0$. (For other incident angles the pictures are similar.) The dashed lines in this figure represent the small $kR$ prediction   (\ref{small_kR}), where we used the following values of $\beta(\alpha)$ computed from the exact formula  (\ref{beta_exact})
\begin{equation}
\beta \left (\frac{1}{4}\right )=-2\ln 2,\;\beta \left (\frac{1}{3} \right )=\ln 2-\frac{3}{2}\ln 3\, ,\; 
\beta\left ( \frac{1}{2}\right )=-\ln 2\ .
\end{equation} 
The agreement between numerics and the asymptotic expression is good,
with differences between them being of the order of $kR$, as expected. 

\section{Exact solution for small $kR$}\label{exact_solution}

The system of equations (\ref{HR}) permits a useful interpretation. 
Let us define the generating functions of $x_n$ and $f_n$
\begin{equation}
\Phi^{(+)}(z)=\sum_{n=0}^{\infty}x_nz^n,
\label{phi_plus}
\end{equation}
and 
\begin{equation}
F(z)=\sum_{n=1}^{\infty}f_n z^n\ 
\end{equation}
which rewrites, using (\ref{f_m}), as
\begin{equation}
F(z)=\frac{\cos \pi \alpha}{4} \left ( \ln (1-z)+ \ln (1+z) \right )
-\frac{\sin \pi \alpha}{2\pi}\left ( \mathrm{Li}_2(z)-\mathrm{Li}_2(-z)\right )
\label{F}
\end{equation}
($\mathrm{Li}_2(z)=\sum_{n=1}^{\infty}z^n/n^2$  is Euler's dilogarithm).
The normalization (\ref{norm}) leads to $\Phi^{(+)}(0)=1$. 

In principle, the $x_n$'s are coefficients of an expansion of a  function holomorphic outside the circle (see Section~\ref{approach}). It would be more appropriate to use instead of  $z=r\mathrm{e}^{\mathrm{i}\phi}$ the variable $1/\bar{z}=\mathrm{e}^{\mathrm{i}\phi}/r$.
To simplify the notations, we use the variable $z$, but retain the superscript $(+)$ in the function (\ref{phi_plus}) to stress that it is related to a  function well behaved outside the circle. 
Equation (\ref{HR}) can then be interpreted as
\begin{equation}
\Phi^{(+)}(\mathrm{e}^{\mathrm{i}\phi})f(\phi)=
\Phi^{(-)}(\mathrm{e}^{\mathrm{i}\phi})+\xi F(\mathrm{e}^{\mathrm{i}\phi})
\label{Hilbert}
\end{equation}
where $\Phi^{(-)}(z)$ has an expansion in non-positive powers of $z$
\begin{equation}
\Phi^{(-)}(z)=\sum_{n=0}^{\infty}\frac{d_n}{z^n}\ .
\end{equation}
Equation (\ref{Hilbert}) is the standard equation for  the non-homogeneous Riemann-Hilbert problem (see, e.g., \cite{singular}), and its solution can be obtained by the usual methods \cite{singular}.
 
First one needs to find the solution of the homogeneous equation
\begin{equation}
T^{(+)}(\mathrm{e}^{\mathrm{i}\phi})f(\phi)=T^{(-)}(\mathrm{e}^{\mathrm{i}\phi})\ .
\label{T}
\end{equation}
which can be done by taking its logarithm 
\begin{equation}
\ln T^{(+)} -\ln T^{(-)} =-\ln f(\phi)\ .
\end{equation}
An elementary solution of the last equation is  the Cauchy integral  \cite{singular}
\begin{equation}
\ln T(z)=-\frac{1}{2\pi}\int_0^{2\pi} \frac{\ln f(\phi)}{1-z\mathrm{e}^{-\mathrm{i}\phi}}\mathrm{d}\phi \ .
\end{equation}
Here for symmetry reasons we use the third definition of $f(\phi)$ in (\ref{phi}) and again drop the factor $\mathrm{e}^{\mathrm{i}\pi\alpha}$, like in (\ref{HR}):
\begin{equation}
f(\phi)=\left \{\begin{array}{cc}
\mathrm{e}^{-\mathrm{i}\pi \alpha},& 0<\phi<\pi\\
\mathrm{e}^{\mathrm{i}\pi \alpha},& \pi <\phi<2\pi\end{array}\right . \ .
\label{new_f}
\end{equation}
It follows that
\begin{equation}
T^{(+)}(z)=\left (\frac{1+z}{1-z}\right )^{\alpha},\;\;T^{(-)}(z)=\left (\frac{1+1/z}{1-1/z}\right )^{\alpha}\ .
\label{T_plus}
\end{equation}
One easily checks that
\begin{equation}
T^{(\pm )}(\mathrm{e}^{\mathrm{i}\phi})= \left |\cot \frac{\phi}{2}\right |^{\alpha}  \left \{\begin{array}{rc}
\mathrm{e}^{\pm  \mathrm{i}\frac{\pi}{2}\alpha} ,& 0<\phi<\pi\\
\mathrm{e}^{\mp \mathrm{i}\frac{\pi}{2}\alpha} ,& \pi <\phi<2\pi \end{array}\right .
\end{equation}
so that  (\ref{T}) is fulfilled. 

Using (\ref{T}), the full non-homogeneous equation (\ref{Hilbert}) can now be rewritten as 
\begin{equation}
\Phi^{(+)}(\mathrm{e}^{\mathrm{i}\phi})\frac{T^{(-)}(\mathrm{e}^{\mathrm{i}\phi})}
{T^{(+)}(\mathrm{e}^{\mathrm{i}\phi})} =
\Phi^{(-)}(\mathrm{e}^{\mathrm{i}\phi})+\xi F(\mathrm{e}^{\mathrm{i}\phi})\ 
\end{equation}
and therefore
\begin{equation}\label{theend}
\frac{\Phi^{(+)}(\mathrm{e}^{\mathrm{i}\phi})}
{T^{(+)}(\mathrm{e}^{\mathrm{i}\phi})} -
\frac{\Phi^{(-)}(\mathrm{e}^{\mathrm{i}\phi})}{T^{(-)}(\mathrm{e}^{\mathrm{i}\phi})}=
\xi \frac{F(\mathrm{e}^{\mathrm{i}\phi})}{T^{(-)}(\mathrm{e}^{\mathrm{i}\phi})}\ .
\end{equation}
A solution of (\ref{theend}) is again given by a  Cauchy integral 
\begin{equation}
\frac{\Phi(z)}{T(z)}=\frac{\xi}{2\pi }\int_0^{2\pi}\frac{F(\mathrm{e}^{\mathrm{i}\phi})}{T^{(-)}(\mathrm{e}^{\mathrm{i}\phi})[1-z\mathrm{e}^{-\mathrm{i}\phi}]}\mathrm{d}\phi \ .
\label{res}
\end{equation} 
Note that  since $F(z)$ and $T^{(-)}(z)$ are regular respectively inside and outside the unit circle, the integral (\ref{res})  cannot be calculated as residues. In the next Section we show that it can be simplified by using a different approach (cf.~(\ref{final_J})). 

A general solution of (\ref{res})  has the form \cite{singular}
\begin{equation}\label{resbis}
\Phi(z)=\xi T(z) J(z)+T(z)P(z)
\end{equation}
where $P(z)$ is a polynomial and 
\begin{equation}
\fl
J(z)=\frac{1}{2\pi }\int_{0}^{\pi} \left [ \mathrm{e}^{\mathrm{i}\pi \alpha/2}\tan^{\alpha}(\phi/2) 
\frac{ F(\mathrm{e}^{\mathrm{i}\phi})}{1-z\mathrm{e}^{-\mathrm{i}\phi}}
+\mathrm{e}^{-\mathrm{i}\pi \alpha/2}\cot^{\alpha}(\phi/2) 
\frac{F(-\mathrm{e}^{\mathrm{i}\phi})}{1+z\mathrm{e}^{-\mathrm{i}\phi}}\right ]\mathrm{d}\phi 
\ .
\label{num}
\end{equation}
Under  $\phi \to \pi-\phi$  (\ref{num}) has the symmetry
\begin{equation}
J(\bar{z})=\bar{J}(z)\ .
\end{equation}
In particular, for $z$ real  $J(z)$ is real and can be written as
\begin{equation}
\fl 
J(z)=\frac{1}{\pi }\int_{0}^{\pi/2}\mathrm{Re}\  \left [ \mathrm{e}^{\mathrm{i}\pi \alpha/2}\tan^{\alpha}(\phi/2) 
\frac{ F(\mathrm{e}^{\mathrm{i}\phi})}{1-z\mathrm{e}^{-\mathrm{i}\phi}}
+\mathrm{e}^{-\mathrm{i}\pi \alpha/2}\cot^{\alpha}(\phi/2) 
\frac{F(-\mathrm{e}^{\mathrm{i}\phi})}{1+z\mathrm{e}^{-\mathrm{i}\phi}}\right ]\mathrm{d}\phi 
\ .
\label{real_z}
\end{equation}
Using $\Phi^{(+)}(0)=1$ and  the fact that $\Phi^{(-)}(z)$ starts at infinity with a constant  we infer from (\ref{resbis}) that 
\begin{equation}
\Phi(z)=\xi T(z)[ J(z)-J(0)]+T(z)\ .
\label{final}
\end{equation}
Clearly (\ref{final}) is a solution of (\ref{Hilbert}) and consequently of (\ref{HR}) for all $\xi$'s. Eq.~(\ref{first_HR}) has yet to  be fulfilled. Formally it can be rewritten as
\begin{equation}
\frac{\pi}{2} \xi \cot \pi \alpha=\lim_{\delta\to 0} [\Phi(1-\delta)-\Phi(-1)]
\label{additional}    
\end{equation}
where $\Phi(z)$ is the same function as in (\ref{phi_plus}) which is given by (\ref{final}). ($\Phi(1)$ in general diverges which explains the necessity of the indicated limit.)

According to (\ref{as_minus}) and (\ref{as_plus}) of Appendix, one has
\begin{equation}
\Phi(-1)=-\frac{\pi}{4} \cot(\pi \alpha)\xi
\label{phi_minus}
\end{equation}
and
\begin{equation}
\Phi(1-\delta)=\left (\frac{2}{\delta}\right )^{\alpha}\left [ \xi(J(1)-J(0))+1 \right ] + \frac{\pi}{4} \cot(\pi \alpha)\xi + \mathcal{O}(\delta^{1-\alpha})\ .
\label{phi_plus_plus}
\end{equation}
Comparing these behaviours with (\ref{additional}), we conclude that $\xi$ has to be chosen in such a way that the singularity of  (\ref{final}) at $z=1$  cancels
\begin{equation}
\xi(J(1)-J(0))+1=0\ .
\label{finite}
\end{equation}
It means that the required  solution of (\ref{HR}) is 
\begin{equation}
\Phi(z)= T(z)\frac{J(1)-J(z)}{J(1)-J(0)}
\label{exact_S_wave}
\end{equation}
where $T(z)$ and $J(z)$ are defined in (\ref{T_plus}) and (\ref{num}) respectively. Expanding $\Phi(z)$ into power of $z$  gives the solution of (\ref{HR}). 

One can now calculate all quantities of interest, in particular, $\beta(\alpha)$ in (\ref{beta})
\begin{equation}
\beta(\alpha)=2(J(1)-J(0))\ ,
\end{equation}
so that using (\ref{real_z}) one finally gets
\begin{equation}
\fl 
\beta(\alpha)=\frac{2}{\pi}\int_0^{\pi/2}\mathrm{Re}\  \left [ \mathrm{e}^{\mathrm{i}\pi \alpha/2}\tan^{\alpha}(\phi/2) 
\frac{ F(\mathrm{e}^{\mathrm{i}\phi})}{\mathrm{e}^{\mathrm{i}\phi}-1}-
\mathrm{e}^{-\mathrm{i}\pi \alpha/2}\cot^{\alpha}(\phi/2) 
\frac{F(-\mathrm{e}^{\mathrm{i}\phi})}{\mathrm{e}^{\mathrm{i}\phi}+1}\right ]\mathrm{d}\phi\ . 
\label{exact_beta}
\end{equation} 
Note that under  $\alpha\to 1-\alpha$, $F(z)\to -F(-z)$,  the integrand in the last expression  remains symmetric and therefore
\begin{equation}
\beta(1-\alpha)=\beta(\alpha)\ ,
\end{equation}
as it follows from (\ref{sym_2}). 

Using (\ref{final_J}), one can prove that when $0<\alpha<1$  (\ref{exact_beta}) is equal to
\begin{equation}
\beta(\alpha)=\frac{1}{2}\Psi\Big (\frac{\alpha}{2}\Big )+\frac{1}{2}\Psi\Big (\frac{1-\alpha}{2}\Big )
+\gamma+\ln 4 +\frac{\pi}{2\sin \pi \alpha}\ .
\label{beta_exact}
\end{equation}
Here the function $\Psi(z)$ is the logarithmic derivative of the gamma function and $\gamma\equiv -\Psi(1)$ is the Euler constant.  $\beta(\alpha)$ calculated numerically from this expression is presented in figure~\ref{fig_alpha}. 

\section{Riemann-Hilbert approach}\label{approach}

In the previous Sections we started from  a system of equations (\ref{main}) valid for small $x=kR$ and  derived an exact solution using the Riemann-Hilbert method which matches two holomorphic functions regular inside and outside a given contour.  Here we demonstrate that the same solution naturally  arises from general considerations without  explicit transformations of boundary equations. 

When $kR\to 0$, one can ignore in the vicinity of the vortices the  $k^2$ term in the Schr\"{o}dinger equation (\ref{schrodinger}) so that the wave function  obeys a Laplace equation. It means that it is a sum of a  function of the variable $z=r\mathrm{e}^{\mathrm{i}\phi}$ and a  function of the variable $\bar{z}=r\mathrm{e}^{-\mathrm{i}\phi}$. Due to the symmetry (\ref{zero_symmetry}), the $S$-wave function should necessarily have the form
\begin{equation}
\Psi(r,\phi)=\tilde{\Phi}(z)+\tilde{\Phi}(-\bar{z})\ .
\label{anti_analytic}
\end{equation}  
Inside the circle  the function $\tilde{\Phi}(z)$ has to be regular and can be expanded into a series in $z$
\begin{equation}
\tilde{\Phi}^{(\mathrm{inside})}(z)=\sum_{m=0}^{\infty}a_m \left (\frac{z}{R}\right )^m .
\end{equation}
Outside the circle the situation is more complicated. The expansion (\ref{elementary}) with the definition (\ref{definition}) states that the outside function has the form
\begin{equation}
\Psi^{(+)}_l(kr,\phi)=J_{|l|}(kr)\mathrm{e}^{\mathrm{i}l\phi}+J_{|l|}(kr)\sum_{n=-\infty}^{\infty}y_n(l) \frac{H_{n}^{(1)}(kr)}{H_{n}^{(1)}(kR)} \mathrm{e}^{\mathrm{i}n\phi}.
\label{element}
\end{equation}
When $kR\to 0$ one gets
\begin{equation}
\Psi^{(+)}_l(kr,\phi)=\tilde{\Phi}^{(+)}(z)+\tilde{\Phi}^{(+)}(-\bar{z})+\rho y_0(0) \ln \frac{r}{R}
\end{equation}
with $\rho$ given in (\ref{rho}). Here
\begin{equation}
\tilde{\Phi}^{(+)}(z)=\frac{1}{2}(1+y_0(0))+\sum_{n=1}^{\infty}y_{-n}(0) \left (\frac{R}{z}\right )^n .
\label{outside_function}
\end{equation}
The logarithmic term appears because of the presence of  $H_0^{(1)}(kr)$ whose  short-distance behaviour  requires to add   $\ln (r/R)$ to the class of admissible functions. $\ln (r/R)$ is neither analytic nor anti-analytic  but  can be written  as a sum of functions with the same symmetry as in (\ref{anti_analytic})  
\begin{equation}
\ln \left (\frac{r}{R}\right ) =\frac{1}{2}\ln \left (-\mathrm{i}\frac{z}{R}\right ) +\frac{1}{2}\ln \left (\mathrm{i}\frac{\bar{z}}{R}\right )\ .
\end{equation}
This argument leads to the conclusion that outside the circle the allowed functions have the form
\begin{equation}
\tilde{\Phi}^{(\mathrm{outside})}(z)=\tilde{\Phi}^{(+)}(z)+\frac{1}{2}\rho y_0(0)\ln \left (-\mathrm{i}\frac{z}{R}\right )\ . 
\label{splitting}
\end{equation}
The function $\tilde{\Phi}^{(+)}(z)$ is regular outside the circle but the logarithmic function is not. In the splitting  (\ref{anti_analytic}) it is implicit that if the function $\tilde{\Phi}(z)$ has a cut at $z=|r|\mathrm{e}^{\mathrm{i}\phi_0}$ with fixed $\phi_0$, it should also have a cut at  $z=|r|\mathrm{e}^{\mathrm{i}(\pi-\phi_0)}$. Therefore one has to define $\ln(-\mathrm{i} z)$ on the unit circle $z=\mathrm{e}^{\mathrm{i}\phi}$  with two cuts, the first one from $1$ to $\infty$ and the second one from  $-\infty$ to $-1$,
\begin{equation}
\frac{1}{2}\ln\left ( -\mathrm{i}\mathrm{e}^{\mathrm{i}\phi}\right )\equiv g(\phi)=\frac{\mathrm{i}}{2}\left \{ \begin{array}{cc} \phi-\frac{\pi}{2} & 0<\phi<\pi \\
\phi-\frac{3\pi}{2} & \pi<\phi<2\pi \end{array}\right . .
\label{g}
\end{equation}
Now the Aharonov-Bohm boundary conditions (\ref{AB_function}) are reduced to the condition that the inside and outside functions of variable $z$ (and of $-\bar{z}$)
are related by 
\begin{equation}
\tilde{\Phi}^{(\mathrm{inside})}(\mathrm{e}^{\mathrm{i}\phi})=f(\phi)\tilde{\Phi}^{(\mathrm{outside})}(\mathrm{e}^{\mathrm{i}\phi})\ .
\label{inside_outside}
\end{equation}
Let us first find  two holomorphic functions 
$F^{(+)}(z)$ and $F^{(-)}(z)$ regular, respectively,  outside and inside the circle such that their difference obey 
\begin{equation}
F^{(-)}(z)-F^{(+)}(z)=f(\phi)g(\phi)
\label{definition_F}
\end{equation} 
with  $f(\phi)$ given in (\ref{phi}) and $g(\phi)$ in (\ref{g}). 

The explicit form of $F(z)$ is  given again by the Cauchy integral \cite{singular}
\begin{equation}
F(z)=\frac{1}{2\pi}\oint \frac{f(\phi)g(\phi)}{1-z\mathrm{e}^{-\mathrm{i}\phi}}\mathrm{d}\phi\ . 
\end{equation}
Using for $f(\phi)$ the  expression (\ref{new_f}) one gets
\begin{equation}
\fl
F(z)=\frac{\mathrm{i}}{4\pi}\left [ \int_0^{\pi} \mathrm{e}^{-\mathrm{i}\pi \alpha}
\frac{\phi-\pi/2}{1-z\mathrm{e}^{-\mathrm{i}\phi}}   \mathrm{d}\phi + 
\int_{\pi}^{2\pi} \mathrm{e}^{\mathrm{i}\pi \alpha}
\frac{\phi-3\pi/2}{1-z\mathrm{e}^{-\mathrm{i}\phi}} \mathrm{d}\phi\right ]=\cos \pi \alpha J_1+\sin \pi \alpha J_2  
\end{equation}
where
\begin{equation}
\fl
J_1=\frac{\mathrm{i}}{4\pi} \int_0^{\pi} (\phi-\pi/2)\left [
\frac{1}{1-z\mathrm{e}^{-\mathrm{i}\phi}}  +
\frac{1}{1+z\mathrm{e}^{-\mathrm{i}\phi}}\right ]  \mathrm{d}\phi=\frac{1}{4}[ \ln(1+z)+\ln(1-z)]
\end{equation}
and 
\begin{equation}
\fl
J_2=\frac{\mathrm{1}}{4\pi} \int_0^{\pi} (\phi-\pi/2)\left [
\frac{1}{1-z\mathrm{e}^{-\mathrm{i}\phi}}  -
\frac{1}{1+z\mathrm{e}^{-\mathrm{i}\phi}}\right ]  \mathrm{d}\phi=
-\frac{1}{2\pi}[ \mathrm{Li}_2(z)-\mathrm{Li}_2(-z)]\ .
\end{equation}
Therefore for $|z|\leq 1$
\begin{equation}
F^{(-)}(z)=\frac{\cos \pi \alpha }{4}[ \ln(1+z)+\ln(1-z)]-\frac{\sin \pi \alpha}{2\pi}[ \mathrm{Li}_2(z)-\mathrm{Li}_2(-z)]
\end{equation}
which coincides with (\ref{F}) and for $|z|\geq 1$
\begin{equation}
F^{(+)}(z)=F^{(-)}(-1/z)\ .
\label{F_plus}
\end{equation}
Knowing $F(z)$ allows to rewrite (\ref{inside_outside}) as
\begin{equation}
\tilde{\Phi}^{(-)}(\mathrm{e}^{\mathrm{i}\phi})=f(\phi)\tilde{\Phi}^{(+)}(\mathrm{e}^{\mathrm{i}\phi})-F^{(+)}(\mathrm{e}^{\mathrm{i}\phi})\rho y_0(0) 
\label{new_hilbert}
\end{equation} 
where
\begin{equation}
\tilde{\Phi}^{(-)}(z)=\tilde{\Phi}^{(\mathrm{inside})}(z)+F^{(-)}(z)\rho y_0(0)
\end{equation}
is a function regular inside the circle and $\tilde{\Phi}^{(+)}(z)$ defined in (\ref{outside_function}) is regular outside the circle. 

To transform this equation to the form solved in the previous Section, we note that 
due to the invariance (\ref{invariance})  $f(\pi-\phi)=f(\phi)$ and (\ref{new_hilbert}) remains valid under $\phi\to \pi-\phi$  
\begin{equation}
\tilde{\Phi}^{(-)}(-\mathrm{e}^{-\mathrm{i}\phi})=f(\phi)\tilde{\Phi}^{(+)}(-\mathrm{e}^{-\mathrm{i}\phi})-F^{(+)}(-\mathrm{e}^{-\mathrm{i}\phi})\rho y_0(0) \ .
\label{transformed_hilbert}
\end{equation}
According to  (\ref{zero_symmetry}) $y_{-n}(0)=(-1)^ny_n(0)$,
so that  
\begin{equation} 
\tilde{\Phi}^{(+)}(-\mathrm{e}^{-\mathrm{i}\phi})=\frac{1}{2}(1+y_0(0))\Phi (\mathrm{e}^{\mathrm{i}\phi})
\end{equation}
with $\Phi(z)$ defined in (\ref{phi_plus}).
Denoting $\tilde{\Phi}^{(-)}(-\mathrm{e}^{-\mathrm{i}\phi})$ by $\Phi^{(-)}(\mathrm{e}^{-\mathrm{i}\phi})$ and taking into account that
\begin{equation}
F^{(+)}(-\mathrm{e}^{-\mathrm{i}\phi})=F^{(-)}(\mathrm{e}^{\mathrm{i}\phi})\equiv F( \mathrm{e}^{\mathrm{i}\phi})
\end{equation}
with $F(z)$ given in (\ref{F}), we conclude that (\ref{transformed_hilbert}) takes the form
\begin{equation}
\Phi^{(+)}(\mathrm{e}^{\mathrm{i}\phi})f(\phi)=
\Phi^{(-)}(\mathrm{e}^{\mathrm{i}\phi})+\xi F(\mathrm{e}^{\mathrm{i}\phi})
\end{equation}
which  coincides with  (\ref{Hilbert}).
 
Imposing the condition that the wave functions remain finite at  both vortices will fix the value of $y_0(0)$ in the same way as above (cf.~(\ref{finite})). Therefore in the end we get the same solution as in the previous Section.

But this approach gives us more than just recalculating  the  solution by a slightly different method. From (\ref{inside_outside}) together with (\ref{splitting})  it is clear that the problem under consideration is equivalent to
\begin{equation}
\fl
\tilde{\Phi}^{(-)}(\mathrm{e}^{\mathrm{i}\phi})=f(\phi)\tilde{\Phi}^{(+)}(\mathrm{e}^{\mathrm{i}\phi})+ 
\frac{\rho }{2}y_0(0)f(\phi)\ln \left (-\mathrm{i}\mathrm{e}^{\mathrm{i}\phi }\right )\equiv 
f(\phi)\tilde{\Phi}^{(+)}(\mathrm{e}^{\mathrm{i}\phi})+ 
\rho y_0(0)f(\phi)g(\phi)
\end{equation}
where the definition (\ref{g}) has been used. 

This equation is also a non-homogeneous Riemann-Hilbert equation of the same type as  (\ref{Hilbert}) with substitution $F(\mathrm{e}^{\mathrm{i}\phi})\longrightarrow f(\phi)g(\phi)$. It means that its solution is the same as above but the integral (\ref{num}) is replaced by 
\begin{equation}
\fl J(z)=\frac{\mathrm{i}}{4\pi }\int_{0}^{\pi} \Big [ \mathrm{e}^{-\mathrm{i}\pi \alpha/2} 
\frac{ \tan^{\alpha}(\phi/2)}{1-z\mathrm{e}^{-\mathrm{i}\phi}}
+\mathrm{e}^{\mathrm{i}\pi \alpha/2} 
\frac{\cot^{\alpha}(\phi/2)}{1+z\mathrm{e}^{-\mathrm{i}\phi}}\Big ]\Big (\phi-\frac{\pi}{2}\Big )\mathrm{d}\phi
\ .
\label{final_J}
\end{equation} 
To prove that this expression and (\ref{num}) are the same we notice that according to (\ref{definition_F}) and (\ref{F_plus}) 
\begin{equation}
F(\mathrm{e}^{\mathrm{i}\phi})-F(-\mathrm{e}^{-\mathrm{i}\phi})=f(\phi)g(\phi)
\end{equation}
so $F(\mathrm{e}^{\mathrm{i}\phi})-f(\phi)g(\phi)$ contains only negative powers of $\mathrm{e}^{\mathrm{i}\phi}$. When this difference is substituted into the integral (\ref{res}) the contour can be shifted to infinity and, therefore, the integrals (\ref{final_J}) are equal to (\ref{num}). Other equivalent forms of this integral will be discussed elsewhere. 

A few general comments about the obtained solution are in order. 
 Combining all terms together, we find that in a vicinity of the vortices but outside the cut circle the $S$-wave function  in the limit $kR\to 0$ takes the form
\begin{equation}
\Psi^{(+)}(\vec{r}\,)=\frac{1}{2}(1+y_0(0))\Big [ \Phi(z)+\Phi(-\bar{z}) \Big ]+ \rho y_0(0)\ln |z| 
\label{main_expansion}
\end{equation}
where $\vec{r}$ has coordinates $x$ and $y$, $z=(x+\mathrm{i}y)/R$, $\bar{z}=(x-\mathrm{i}y)/R$.

In this expression, $\Phi(z)$ is given by (\ref{exact_S_wave}) with $z\to 1/\bar{z}$. The scattering amplitude $y_0(0)$ from the $H_0^{(1)}(kr)$ contribution in the expansion (\ref{elementary}) is fixed by (\ref{beta_alpha}) and (\ref{exact_beta}). That value of $y_0(0)$ renders the function $\Phi(z)$ finite at both vortex locations. In fact, as a consequence of (\ref{phi_minus}) and (\ref{phi_plus_plus}), the total wave function (\ref{main_expansion}) vanishes at the vortex positions, as it is the case for the Aharonov-Bohm effect \cite{AB}
\begin{equation}
\Psi(\vec{r}\,) \sim \left \{\begin{array}{ll}|\vec{r}-\vec{R}|^{\alpha}, & \vec{r}\ \mathrm{close\ to \ the\  right\ vortex\ with\ flux\ \alpha}\\
 |\vec{r}+\vec{R}|^{1-\alpha},&\vec{r}\ \mathrm{close\ to \ the\  left\ vortex\ with\ flux\ -\alpha}\end{array}\right . .
\label{singular_limits}
\end{equation}
In deriving these limits, we take into account that due to the necessary substitution $z\to 1/\bar{z}$, singularities at $z=\pm 1$ in (\ref{exact_S_wave}) and in  (\ref{main_expansion}) are interchanged.  

In this Section we use only one of the two Aharonov-Bohm boundary conditions (\ref{inside_outside}), which matches the function on the cut circle, but do not discuss explicitly the second condition (\ref{AB_derivative}), which matches the derivative. The reason is that the functions we consider are either analytic or anti-analytic.  But for any such function the derivatives with respect to $r$ and $\phi$ are proportional
to each other:
\begin{equation}
\frac{\partial}{\partial r}\Psi(z)=-\frac{\mathrm{i}}{r}\frac{\partial}{\partial \phi}\Psi(z)\ ,\;\;
\frac{\partial}{\partial r}\Psi(\bar{z})=\frac{\mathrm{i}}{r}\frac{\partial}{\partial \phi}\Psi(\bar{z})\ .
\end{equation} 
Therefore, when condition (\ref{inside_outside}) is satisfied and the function $f(\phi)$ is piecewise constant (like in our case) the derivative with respect to $r$ will obey the same condition as the function itself, except for $\delta$-function contributions at the points of discontinuities of $f(\phi)$. One can check that (\ref{first}) is related with the cancellation of these discontinuities. As has been demonstrated in the previous Section, that equation leads to the vanishing of the wave function at singular points (cf.~(\ref{singular_limits})), so the $\delta$-functions give no contributions and the derivatives over $r$ automatically obey the correct boundary condition (\ref{AB_derivative}).  

In \cite{nambu} it was argued that a solution of the $S$-wave type which remains finite at the two vortex locations cannot  exist. But a logarithmic term like the one in (\ref{main_expansion}) was not properly taken into account in that discussion. When  ignored  (which means setting $\xi=0$ in (\ref{exact_S_wave})), only the singular solution $T(z)$ is obtained. The role of the logarithmic term is precisely to remove the singularity at the vortex locations.

\section{Conclusion}\label{conclusion}
We have considered the  scattering problem on two Aharonov-Bohm vortices with opposite fluxes $\alpha$ and $-\alpha$, separated by a distance $2R$.  
First, we developed a numerical method for the construction of scattering amplitude. In its simplest formulation it reduces to solving an infinite system of equations~(\ref{rescaled}). The convergence of finite approximants is slow due to the singular behaviour  of the wave function at the vortex positions (cf.~(\ref{singular_limits})).  
Then we analytically constructed the dominant contributions to the scattering amplitude in two particular  cases:
small values of $\alpha$  and  small values of $kR$. 

In the small $\alpha$ limit, i.e., when $\sin \pi \alpha\to 0$, the scattering amplitude takes the form 
\begin{equation} 
F(\theta,\theta^{\prime})\underset{\sin \pi \alpha\to 0}{\longrightarrow}
-\mathrm{i}\sin \pi \alpha \ \cot \frac{\theta-\theta^{\prime}}{2}\ \sin \left (2x\sin \frac{\theta-\theta^{\prime}}{2}\sin \frac{\theta+\theta^{\prime}}{2}  \right )
\end{equation}
where $\theta$ and $\theta^{\prime}$ are respectively the incident  and  reflection  angles. 

In the  small  $kR$ limit, the solution has been obtained by the Riemann-Hilbert method in its simplest setting. The dominant contribution  is given by the $S$-wave  scattering amplitude, which has the form 
\begin{equation}
F(\theta,\theta^{\prime})\underset{kR\to 0}{\longrightarrow} -\frac{\pi}{\pi+2\mathrm{i}(\ln (kR/2)+\gamma+\beta(\alpha))}
\end{equation}
where $\beta(\alpha)$ is  
\begin{equation}
\beta(\alpha)=\frac{1}{2}\Psi\Big (\frac{\alpha}{2}\Big )+\frac{1}{2}\Psi\Big  (\frac{1-\alpha}{2}\Big )
+\gamma+\ln 4 +\frac{\pi}{2\sin \pi \alpha}
\end{equation}
with $\Psi(z)$ being the logarithmic derivative of the gamma function. 

The same method permits also to find the behaviour of the  $S$-wave function  in the vicinity of the vortices. Denoting this wave function as 
\begin{equation}
\Psi^{(+)}_0(r,\phi)=J_0(kr)+J_0(kR)\sum_{n=-\infty}^{\infty}\frac{y_n(0)}{H_n^{(1)}(kR)}H_n^{(1)}(kr) \mathrm{e}^{\mathrm{i}n\phi}
\end{equation}
we found that when $kR\to 0$
\begin{equation}
y_0(0)=-\frac{1}{1+\beta(\alpha)\rho}
\end{equation}
where 
\begin{equation}
\rho=\frac{2\mathrm{i}}{\pi+2\mathrm{i}(\ln (kR/2)+\gamma)}\ .
\end{equation}
The generating function of $y_n(0)=y_{-n}(0)$ in this limit takes the form
\begin{equation}
\sum_{n=0}^{\infty}y_n(0)z^n=\frac{1+y_0}{2}\Phi(z)-\frac{1}{2}\ .
\end{equation}
Here
\begin{equation}
\Phi(z)= \frac{2}{\beta(\alpha)}\left ( \frac{1+z}{1-z}\right )^{\alpha}[J(1)-J(z)]
\end{equation}
and 
\begin{equation}
J(z)=\frac{\mathrm{i}}{4\pi }\int_{0}^{\pi} \Big [ \mathrm{e}^{-\mathrm{i}\pi \alpha/2} 
\frac{ \tan^{\alpha}(\phi/2)}{1-z\mathrm{e}^{-\mathrm{i}\phi}}
+\mathrm{e}^{\mathrm{i}\pi \alpha/2} 
\frac{\cot^{\alpha}(\phi/2)}{1+z\mathrm{e}^{-\mathrm{i}\phi}}\Big ]\Big (\phi-\frac{\pi}{2}\Big )\mathrm{d}\phi 
\ .
\end{equation}

A discussion of other partial waves in the limit $kR\to 0$ and of related questions in the general case of two Aharonov-Bohm vortices with arbitrary fluxes will be given elsewhere. 
The exact solution for the scattering amplitude  with  $kR$ finite remains an open challenge. 

\appendix

\section{Limiting values}

To calculate the behaviour of the solution (\ref{final}) at $z\to 1$ and $z\to -1$, one has to investigate the corresponding limits for the integral (\ref{real_z}). When $z=-1+\delta$ and $\delta \to 0$, the second term in (\ref{num}) is formally diverging. This divergence appears at small $\phi$ and  can be isolated by splitting the integral  (\ref{real_z}) into two parts:
\begin{eqnarray}
J(-1+\delta)&\equiv &\frac{1}{\pi }\int_{0}^{\pi/2} \mathrm{Re}\left [ \mathrm{e}^{\mathrm{i}\pi \alpha/2}\tan^{\alpha}(\phi/2) 
\frac{ F(\mathrm{e}^{\mathrm{i}\phi})}{1-(-1+\delta)\mathrm{e}^{-\mathrm{i}\phi}}\right .\nonumber\\
&+&\left . \mathrm{e}^{-\mathrm{i}\pi \alpha/2}\cot^{\alpha}(\phi/2) 
\frac{F(-\mathrm{e}^{\mathrm{i}\phi})}{1+(-1+\delta)\mathrm{e}^{-\mathrm{i}\phi}}\right ]\mathrm{d}\phi 
\nonumber \\
&=&\frac{1}{\pi }\int_{0}^{\pi/2}\mathrm{Re} \left [ \mathrm{e}^{-\mathrm{i}\pi \alpha/2}\cot^{\alpha}(\phi/2) 
\frac{F(-\mathrm{e}^{\mathrm{i}\phi})}{\mathrm{i}(\phi -\mathrm{i}\delta)}\right ] \mathrm{d}\phi +\mathcal{O}(1)\ .
\end{eqnarray}
Changing the variable in the last integral $\phi\to t \delta $ gives
\begin{equation}
J(-1+\delta)\underset{\delta\to 0}{\longrightarrow} W(\delta)+\mathcal{O}(1)
\end{equation}
where
\begin{equation}
W(\delta)=\frac{1}{\pi }\int_{0}^{\pi/(2\delta)}\mathrm{Re}\left [  \mathrm{e}^{-\mathrm{i}\pi \alpha/2}\cot^{\alpha}(\delta t /2) 
\frac{F(-\mathrm{e}^{\mathrm{i}\delta t})}{\mathrm{i}(t -\mathrm{i})}\right ]\mathrm{d}t\ . 
\end{equation}
From known expressions 
\begin{equation}
\mathrm{Li}_2(1)=\frac{\pi^2}{6}\ ,\;\;\mathrm{Li}_2(-1)=-\frac{\pi^2}{12}
\end{equation}
it follows that the function $F(-\mathrm{e}^{\mathrm{i}\phi})$ has the following expansion   
\begin{equation}
F(-\mathrm{e}^{\mathrm{i}\phi})\underset{\phi \to 0}{\longrightarrow} \frac{1}{4}\cos \pi \alpha [\ln 2 \phi -\mathrm{i}\frac{\pi}{2}]+\frac{\pi}{8}\sin \pi \alpha \ .
\label{ser_F_minus}
\end{equation}
Finally, one has in the limit $\delta\to 0$
\begin{equation}
\fl
W(\delta)=\left (\frac{2}{\delta}\right )^{\alpha}\mathrm{Re }  \int_{0}^{\infty} \frac{\mathrm{e}^{-\mathrm{i}\pi \alpha/2}} 
{\mathrm{i}\pi t^{\alpha}(t -\mathrm{i})} \left [ \frac{1}{4}\cos \pi \alpha [\ln( 2 \delta t) +\mathrm{i}\frac{\pi}{2}]+\frac{\pi}{8}\sin \pi \alpha \right ]\mathrm{d}t\ .
\label{W}
\end{equation}
The first integral in this expression can be calculated by contour integration
\begin{equation}
\int_{0}^{\infty}\frac{\mathrm{d}t}{t^{\alpha}(t -\mathrm{i})}=\frac{\pi}{\sin \pi \alpha}\mathrm{e}^{\mathrm{i}\pi \alpha/2}
\end{equation}
and the second one by taking the derivative with respect to $\alpha$ of this result
\begin{equation}
\int_{0}^{\infty}\frac{\mathrm{d}t\ln t}{t^{\alpha}(t -\mathrm{i})}=\frac{\pi^2}{\sin \pi \alpha}\mathrm{e}^{\mathrm{i}\pi \alpha/2}(-\frac{\mathrm{i}}{2}+\cot \pi\alpha)\ .
\end{equation}
Substituting these expressions into (\ref{W}) and taking the real part, one obtains
\begin{equation}
W(\delta)=-\left (\frac{2}{\delta}\right )^{\alpha}\frac{\pi \cot(\pi \alpha)}{4} 
\end{equation} 
and 
\begin{equation}
J(-1+\delta)\underset{\delta\to 0}{\longrightarrow} -\left (\frac{2}{\delta}\right )^{\alpha}\frac{\pi \cot(\pi \alpha)}{4}+   \mathcal{O}(1)\ .
\end{equation}
Because $T(-1+\delta)\to (\delta/2)^{\alpha}$, from (\ref{final}) it follows that
\begin{equation}
\Phi(-1+\delta)=-\frac{\pi}{4} \cot(\pi \alpha)\xi + \mathcal{O}(\delta^{\alpha})\ .
\label{as_minus}
\end{equation}
The behaviour of $J(z)$ close to $z=1$ can be obtained by a similar method. The difference is that in this case the value $J(1)$ is finite but due to the singularity of $T(z)$ at $z=1$ we also need to know the correction term. One has 
\begin{eqnarray}
J(1-\delta)&=& \frac{1}{\pi } \mathrm{Re}\int_{0}^{\pi/2}\left [ \mathrm{e}^{\mathrm{i}\pi \alpha/2}\tan^{\alpha}(\phi/2) 
\frac{ F(\mathrm{e}^{\mathrm{i}\phi})}{1-(1-\delta)\mathrm{e}^{-\mathrm{i}\phi}}\right . \nonumber\\
 &+&\left . \mathrm{e}^{-\mathrm{i}\pi \alpha/2}\cot^{\alpha}(\phi/2) 
\frac{F(-\mathrm{e}^{\mathrm{i}\phi})}{1+(1-\delta)\mathrm{e}^{-\mathrm{i}\phi}}\right ]\mathrm{d}\phi 
\end{eqnarray}
First  we split this integral  into two parts as follows
\begin{equation}
\int_0^{\pi/2}\ldots \mathrm{d}\phi=
\int_0^{\epsilon}\ldots \mathrm{d}\phi+
\int_{\epsilon}^{\pi/2}\ldots \mathrm{d}\phi
\label{split}
\end{equation}
where $\delta \ll \epsilon\ll 1$. 

In the first integral, the first term dominates at small $\phi$, and 
\begin{equation}
J(1-\delta)\underset{\delta\to 0}{\longrightarrow} J(1)+V(\delta)
\end{equation}
where
\begin{equation}
V(\delta)=\frac{1}{\pi }\mathrm{Re}\ \int_{0}^{\epsilon}  \mathrm{e}^{\mathrm{i}\pi \alpha/2}\tan^{\alpha}(\phi/2) 
F(\mathrm{e}^{\mathrm{i}\phi})
\left (\frac{1}{\mathrm{i}(\phi-\mathrm{i}\delta)} -\frac{1}{\mathrm{i}\phi}\right ) \mathrm{d}\phi \ .
\end{equation}  
The second term in the parentheses appears when the splitting (\ref{split})
is applied for the calculation of $J(1)$. 

Using an expansion similar to (\ref{ser_F_minus})
\begin{equation}
F(\mathrm{e}^{\mathrm{i}\phi})\underset{\phi \to 0}{\longrightarrow} \frac{1}{4}\cos \pi \alpha [\ln 2 \phi -\mathrm{i}\frac{\pi}{2}]-\frac{\pi}{8}\sin \pi \alpha 
\label{ser_F_plus}
\end{equation}
and rescaling the variable $\phi=t\delta$, one concludes that   
\begin{equation}
\fl
V(\delta)=\left (\frac{\delta}{2}\right )^{\alpha}
\mathrm{Re}\ \int_{0}^{\infty}  \mathrm{e}^{\mathrm{i}\pi \alpha/2} \frac{t^{\alpha}}{\pi t(t-\mathrm{i})} 
\left [\frac{1}{4}\cos \pi \alpha [\ln (2 t \delta)  -\mathrm{i}\frac{\pi}{2}]-\frac{\pi}{8}\sin \pi \alpha  \right ]\mathrm{d}t\ .
\end{equation} 
The remaining integrals are calculated as above:
\begin{equation}
\fl \int_{0}^{\infty} \frac{t^{\alpha}}{t(t-\mathrm{i})}\mathrm{d}t=\frac{ \mathrm{i}\pi}{\sin \pi \alpha}\mathrm{e}^{-\mathrm{i}\pi \alpha/2} 
\ , \;
\int_{0}^{\infty} \frac{t^{\alpha}\ln t}{t(t-\mathrm{i})}\mathrm{d}t=\frac{\mathrm{i}\pi^2 }{\sin \pi \alpha}\mathrm{e}^{-\mathrm{i}\pi \alpha/2}\left ( -\frac{\mathrm{i}}{2} -\cot \pi \alpha \right ). 
\end{equation}
Using these values, one gets
\begin{equation}
V(\delta)=\left (\frac{\delta}{2}\right )^{\alpha}\frac{\pi}{4}\cot \pi \alpha\ .
\end{equation} 
As $T(1-\delta) \underset{\delta \to 0}{\longrightarrow} (2/\delta)^{\alpha}$
we conclude that
\begin{equation}
\Phi(1-\delta)=\left (\frac{2}{\delta}\right )^{\alpha}\left [ \xi(J(1)-J(0))+1 \right ] + \frac{\pi}{4} \cot(\pi \alpha)\xi + \mathcal{O}(\delta^{1-\alpha})\ . 
\label{as_plus}
\end{equation}

\end{document}